\documentclass[11pt,a4paper]{article}
\usepackage{jheppub}
\usepackage{amsmath,amssymb}
\usepackage{graphicx}
\usepackage{dcolumn}
\usepackage{bm}
\usepackage{slashed}
\usepackage{setspace} 
\usepackage{ulem}
\usepackage{braket}
\usepackage{color}
\usepackage{subfigure}
\usepackage{float}
\usepackage{amsfonts}
\usepackage{booktabs}
\usepackage{caption}
\usepackage{enumitem}
\begin{document}

\title{Beyond the Standard Model Higgs physics: Hunting $h\to bs$ with Higgs-strahlung at CEPC and FCC-ee}

\author[a,b]{M. A. Arroyo-Ure\~na}
\author[a,b]{D. A. Carre\~no-Diaz}
\author[a,b]{A. I. Garc\'ia-Gutierrez}
\author[c]{T.A. Valencia-P\'erez}
\author[a,b]{M. G. Villanueva-Utrilla}

\affiliation[a]{Facultad de Ciencias F\'isico-Matem\'aticas, Benem\'erita Universidad Aut\'onoma de Puebla, C.P. 72570, Puebla, M\'exico}
\affiliation[b]{Centro Interdisciplinario de Investigaci\'on y Ense\~nanza de la Ciencia (CIIEC), Benem\'erita Universidad Aut\'onoma de Puebla, C.P. 72570, Puebla, M\'exico}
\affiliation[c]{Instituto de F\'isica, Universidad Nacional Aut\'onoma de M\'exico, C.P. 01000, CDMX, M\'exico}

\emailAdd{marco.arroyo@fcfm.buap.mx}
\emailAdd{diego.carrenod@alumno.buap.mx}
\emailAdd{alejandro.garciagut@alumno.buap.mx}
\emailAdd{tvalencia@fisica.unam.mx}
\emailAdd{maria.villanueva@alumno.buap.mx}

\maketitle
\begin{abstract}

The next generation of circular electron-positron colliders, such as the Circular Electron Positron Collider and the Future Circular Collider, will provide an unprecedented level of precision in probing the properties of the Higgs boson. In this study, we examine the flavor-changing neutral Higgs decay $h\to bs$ produced via the Higgs-strahlung process within the framework of the Two-Higgs-Doublet Model Type III. We identify regions of the parameter space that remain viable after applying stringent constraints from current experimental data. By employing a multivariate analysis based on Boosted Decision Trees, we project that a signal significance of $5\sigma$ can be achieved, even after accounting for irreducible backgrounds and systematic uncertainties. Our findings show that these future colliders could offer a unique avenue for discovering signatures of physics beyond the standard model.
\end{abstract}



\section{Introduction}\label{sec:Introduction}
In 2012, the ATLAS and CMS collaborations announced the observation of a new spin-0 particle with mass $m_h \approx 125$ GeV~\cite{CMS:2013btf, ATLAS:2012yve} consistent with the Higgs boson properties predicted by the Standard Model (SM).  After its discovery, the Large Hadron Collider (LHC) has characterized the properties of the Higgs boson because, among other things, the Higgs field is fundamental to the generation of the particle masses contained in the SM~\cite{Englert:1964et, Higgs:1964pj}. As is well known, the SM provides a successful description of the weak, strong, electromagnetic interactions, and many of the experimental observations made in particle physics. Despite these achievements, there are phenomena that cannot be explained within the SM. In particular, Flavor-Changing Neutral Currents (FCNC) Higgs decays in the quark sector are highly suppressed due to the GIM mechanism~\cite{Maiani:2013fpa} and the smallness of the Cabibbo-Kobayashi-Maskawa ($V_{\rm CKM}$) quark mixing matrix elements. These decays involve the Higgs boson transforming into a different quark flavor $h\to q_iq_j$. The theoretical SM predictions of the different channels are given by $\mathcal{BR}(h\to bs)=1.78\times10^{-7}$, $\mathcal{BR}(h\to bd)=8.36\times10^{-9}$, $\mathcal{BR}(h\to sd)=8.65\times10^{-15}$, $\mathcal{BR}(h\to cu)=8.13\times10^{-19}$~\cite{Farrera:2020bon}. Since the SM predictions are extremely suppressed, observe them at different rates at colliders like the LHC would be a strong indication of physics beyond the SM. However, analyze decays like $h\to q_i q_j$ at proton-proton colliders represent serious challenges, highlighting the large background of QCD which makes it very difficult to identify these kind of processes. In contrast, the next generation of electron-positron colliders, complementary to the LHC, can study the properties of the Higgs boson -and other particles- with unprecedented precision, high integrated luminosity, high center-of-mass energies ---relative to previous electron-positron colliders like LEP \cite{Schopper:2009zz}, which reached $\sqrt{s} \sim 209$ GeV---, within a very clean background environment, making them an excellent option for copiously studying FCNC Higgs decays. Unlike to the Higgs boson production in proton-proton collisions, the environment of $e^-e^+$ collisions will significantly reduce SM background processes that obscure the signals studied at the LHC or future stages of it, namely, the High-Luminosity LHC~\cite{Apollinari:2015wtw}, High-Energy LHC~\cite{Cepeda:2019klc}, and the Future hadron-hadron Circular Collider~\cite{Arkani-Hamed:2015vfh}. In this context, the particle physics community has been engaged for many years in the design of several $e^-e^+$ collider projects: Circular Electron-Positron Collider (CEPC)~\cite{CEPCStudyGroup:2023quu}, Future $ee$ Circular Collider (FCC-ee)~\cite{FCC:2018evy}, International Linear Collider (ILC)~\cite{ILCInternationalDevelopmentTeam:2022izu}, and the Compact LInear Collider (CLIC)~\cite{Adli:2025swq}. In this work, we focus on the CEPC and FCC-ee colliders because Higgs boson production -via  the Higgs-strahlung process- is highly favored at a center-of-mass energy of $\sqrt{s}=240$ GeV. This energy, combined with their projected high integrated luminosity (up to several ab$^{-1}$) makes CEPC and FCC-ee more advantageous for precision Higgs studies compared to linear colliders like the ILC and CLIC. While the latter can reach higher energies, their integrated luminosities are expected to be significantly lower than those achievable by circular colliders at $240$ GeV. The branching ratios for the decays $h\to bs$ and $h\to uc$ can be measured at the CEPC with an upper limit $\sim 0.03\%$ and $0.08\%$ at $95\%$ C.L.~\cite{Ai:2024nmn}, respectively. A study at the FCC-ee~\cite{Kamenik:2023hvi} indicates comparable sensitivities of measuring the rates of $h\to bs$ and $h\to uc$, estimating the upper limits to be $\mathcal{O}(10^{-3})$. With the possible arrival of these colliders, which have as one of their priorities the study of the Higgs boson, non-standard phenomena could be brought to experimental scrutiny. Due to the advantages of $e^-e^+$ colliders, FCNC could be of special interest, as they are predicted to have sizeable branching ratios within the theoretical framework of SM extensions. A well-motivated model is the Two-Higgs Doublet Model of type III (2HDM-III), which induces FCNC at tree-level and predicts large rates which can be simultaneously controlled by several mechanisms, such as four-zero textures~\cite{Fritzsch:1995nx, Branco:1999nb, LorenzoDiaz-Cruz:2019imm, Diaz-Cruz:2004wsi, Arroyo-Urena:2013cyf}, which are invoked in this work.

Thus, we are interested in exploring the prospects of detecting the flavor-changing $h\to b\bar{s}+\bar{b}s$ decay via the Higgs-strahlung production at CEPC and FCC-ee in the 2HDM-III as it predicts branching ratios for the decay $h\to bs$ of up to $\mathcal{O}(10^{-3})$.

	This work is structured as follows. In Sec. \ref{SecII}, we describe the relevant aspects of the theoretical framework adopted in this work, $\textit{i.e.}$, the 2HDM-III with a particular emphasis on the theoretical implications of employing a four-zero texture in the Yukawa Lagrangian. An analysis of the model parameter space is also included. Section \ref{SecIII} is focused on the study of the proposed signal and its SM background processes. Finally, the conclusions are presented in Sec. \ref{SecV}.

\section{Two-Higgs Doublet Model of type III.}\label{SecII}


The Two-Higgs Doublet Model includes an additional scalar doublet in addition to the SM. We denote these doublets as $\Phi_i^T=( \phi_{i}^{+},\phi_{i}^{0})$ ($i=1, 2$). After Spontaneous Symmetry Breaking (SSB), the Higgs doublets acquire nonzero vacuum expectation values (VEV), given as $\braket{\Phi_i}^T=1/\sqrt{2}\,(0\;\;\upsilon_i)$. {The SM Higgs VEV, \( v \approx 246 \) GeV, is related to these VEVs by \( v^2 = v_1^2 + v_2^2 \).}
	
	The introduction of additional scalar doublet predicts four particles (two neutral and two charged), and gives place to new interactions between the doublets and fermions. The 2HDM-III also leads Flavor-Changing Neutral Interactions (FCNI) at the tree level, which must to be suppressed in order to meet with the experimental observations. In this work, we assume a specific structure of the Yukawa matrices $Y_i$, the so-called four-zero texture.{This means that in a chosen weak basis, exactly four independent entries of each matrix are set to zero, while the remaining non-zero elements are generally complex. This constrained pattern leads to predictive relations among fermion masses and the $V_{\rm CKM}$ matrix elements. Such Hermitian four-zero texture matrices are well-motivated in the literature~\cite{Fritzsch:1995nx, Branco:1999nb, LorenzoDiaz-Cruz:2019imm, Diaz-Cruz:2004wsi, Arroyo-Urena:2013cyf}}. Using it, we can effectively control the strength of FCNIs and ensure consistency with the experimental data.
    
\subsection{Scalar potential}
The most general $SU(2)_L \times U(1)_Y$ invariant scalar potential is given by:  

 \begin{eqnarray}
V(\Phi_1,\Phi_2) &=& \mu_{1}^{2}(\Phi_{1}^{\dag}\Phi_{1}^{}) + \mu_{2}^{2}(\Phi_{2}^{\dag}\Phi_{2}^{}) - \mu_{12}^{2}(\Phi_{1}^{\dag}\Phi_{2}^{} + h.c.) \nonumber \\ 
&+& \frac{1}{2} \lambda_{1}(\Phi_{1}^{\dag}\Phi_{1}^{})^2 + \frac{1}{2} \lambda_{2}(\Phi_{2}^{\dag}\Phi_{2}^{})^2 \nonumber \\ 
&+& \lambda_{3}(\Phi_{1}^{\dag}\Phi_{1}^{})(\Phi_{2}^{\dag}\Phi_{2}^{}) + \lambda_{4}(\Phi_{1}^{\dag}\Phi_{2}^{})(\Phi_{2}^{\dag}\Phi_{1}^{}) \nonumber \\
&+& \Big[\frac{1}{2} \lambda_{5}(\Phi_{1}^{\dag}\Phi_{2}^{})^2  + \lambda_{6}(\Phi_{1}^{\dag}\Phi_{1}^{})(\Phi_{1}^{\dag}\Phi_{2}^{}) \nonumber \\
&+& \lambda_{7}(\Phi_{2}^{\dag}\Phi_{2}^{})(\Phi_{1}^{\dag}\Phi_{2}^{}) + h.c.\Big].
\label{potential}
\end{eqnarray}

We assume that the potential parameters $\lambda_{5,\,6,\,7}$, $\mu_{12}$ and the VEVs are real. Thus, the CP symmetry is conserved. After the diagonalization of the scalar potential in Eq.~\eqref{potential}, the masses of the new particles predicted in the model are given by:
\begin{eqnarray}
	M_{H^{\pm}}^2&=&\frac{\mu_{12}^2}{\sin\beta\cos\beta}-\frac{1}{2}v^2\Big(\lambda_4+\lambda_5+\cot\beta\lambda_6+\tan\beta\lambda_7\Big)\nonumber \\
	M_{A}^2&=&M_{H^{\pm}}^2+\frac{1}{2}v^2(\lambda_4-\lambda_5),\nonumber\\
	M_{H,\,h}^2&=&\frac{1}{2}\Big(m_{11}+m_{22}\pm \sqrt{(m_{11}-m_{22})^2+4m_{12}^2}\Big),
\end{eqnarray}
where $m_{ij}$ corresponds to the elements of the real part of the mass matrix $\mathbf{M}$,
	\begin{center}
		$\rm Re(\bold{M})=\left(\begin{array}{cc}
			m_{11} & m_{12}\\
			m_{12} & m_{22}
		\end{array}\right)$,
		\par\end{center}
and are given by
    \begin{eqnarray}
		m_{11}&=&v^2(\lambda_1\cos^2\beta+\lambda_{5}\sin^2\beta)+\sin^2\beta\, M_A^2,\nonumber
		\\
		m_{12}&=&v^2(\lambda_{3}+\lambda_{4})\cos\beta\sin\beta-\cos\beta\sin\beta M_A^2,\nonumber 
		\\
		m_{22}&=&v^2(\lambda_{2}\sin^2\beta+\lambda_{5}\cos^2\beta)+\cos^2\beta\, M_A^2.
	\end{eqnarray}
The angle $\beta$ defines the VEVs ratio as $\tan\beta=\upsilon_2/\upsilon_1$. Such an angle mixes the imaginary part with the neutral component of $\Phi_i$ as well as their charged components, as follows:
\begin{eqnarray}
	G_W^{\pm}&=&\phi_1^{\pm}\cos\beta+\phi_2^{\pm}\sin\beta,\nonumber\\
		H^{\pm}&=&-\phi_1^{\pm}\sin\beta+\phi_2^{\pm}\cos\beta,\nonumber\\	
		G_Z&=&\rm Im(\phi_1^0)\cos\beta+\rm Im(\phi_2^0)\sin\beta,\nonumber\\
		A&=&-\rm Im(\phi_1^0)\sin\beta+\rm Im(\phi_2^0)\cos\beta,
	\end{eqnarray}
from which we obtain the charged scalar bosons $H^{\pm}$, the pseudo-Goldstone boson associated with the $W$ gauge fields, a CP-odd state $A$, and the pseudo-Goldstone boson related to the $Z$ gauge boson.
Meanwhile, the real part of the scalar doublets $\Phi_i$ define two CP-even states through the mixing angle $\alpha$:	
	\begin{eqnarray}
		H&=&\rm Re(\phi_1^0)\cos\alpha+\rm Re(\phi_2^0)\sin\alpha,\nonumber\\
		h&=&\rm Re(\phi_1^0)\sin\alpha+\rm Re (\phi_2^0)\cos\alpha,
		\label{eq:Hh_definition}
	\end{eqnarray}
	where
	\begin{equation*}
		\tan2\alpha=\frac{2m_{12}}{m_{11}-m_{22}}.
	\end{equation*}	
Eq.~\eqref{eq:Hh_definition} gives rise to the SM-like Higgs boson and a CP-even scalar $H$.
\subsection{Yukawa Lagrangian}
The Yukawa Lagrangian of the 2HDM-III reads \cite{HernandezSanchez:2012eg}:
	\begin{align}
\label{YukawaLagrangian} 
\mathcal{L}_Y = & -\left( Y_{1}^{u} \bar{Q}_{L} \tilde{\Phi}_{1} u_{R} + Y_{2}^{u} \bar{Q}_{L} \tilde{\Phi}_{2} u_{R} \right. \nonumber \\
& \left. + Y_{1}^{d} \bar{Q}_{L} \Phi_{1} d_{R} + Y_{2}^{d} \bar{Q}_{L} \Phi_{2} d_{R} \right. \nonumber \\
& \left. + Y_{1}^{l} \bar{L}_{L} \tilde{\Phi}_{1} l_{R} + Y_{2}^{l} \bar{L}_{L} \tilde{\Phi}_{2} l_{R} \right),
\end{align}
where  $\tilde{\Phi}_{j} = i\sigma_2 \Phi_{j}^{*}$ $(j=1, 2)$.  After the SSB, the fermion mass matrices can be defined as:
	
	\begin{equation}
		M_f = \frac{1}{\sqrt{2}} \left( v_1 Y_{1}^{f} + v_2 Y_{2}^{f} \right),\; f=u,d,\ell.	
	\end{equation}
The Yukawa matrices are assumed to exhibit a specific \textit{four-zero texture} structure; in our framework, they take the form:
\begin{equation}
Y_{1}^{f}=\left(\begin{array}{ccc}
	0 & d_{1}^{f} & 0\\
	d_{1}^{f} & c_{1}^{f} & b_{1}^{f}\\
	0 & b_{1}^{f} & a_{1}^{f}
\end{array}\right),\qquad Y_{2}^{f}=\left(\begin{array}{ccc}
	0 & d_{2}^{f} & 0\\
	d_{2}^{f} & c_{2}^{f} & b_{2}^{f}\\
	0 & b_{2}^{f} & a_{2}^{f}
\end{array}\right).
\end{equation}
The sum of the Yukawa matrices inherits its form to the mass matrix as follows
\begin{equation}\label{4zero_MASS_matrices}
	M_f=\left(\begin{array}{ccc}0 & D_f & 0 \\D_f & C_f & B_f \\0 & B_f & A_f\end{array}\right).\;
\end{equation}
The elements of the fermion mass matrices in Eq.~\eqref{4zero_MASS_matrices}
are related to the physical fermion masses $m_{f_i}$ ($i=1,\,2,\,3$), via the following invariants:
\begin{equation}
	\begin{array}{rcl}
		\text{Tr}\left(M_f\right) & = & C_f+A_f=m_{f_1}+m_{f_2}+m_{f_3},\\
		\lambda\left(M_f\right) & = & C_fA_f-D_f^{2}-B_f \\&=&m_{f_1}m_{f_2}+m_{f_1}m_{f_3}+m_{f_2}m_{f_3},\\ \text{det}\left(M_f\right) & = & -D_f^{2}A_f=m_{f_1}m_{f_2}m_{f_3}.
		\label{Invariantes} 
	\end{array}
\end{equation}
From Eq.~\eqref{Invariantes}, we obtain the relation between the mass matrix elements in its four-zero texture structure and the fermion masses as follows:
\begin{eqnarray}\label{ElementosMatrizMasa}
	A_f & = & m_{f_3}-m_{f_2},\nonumber \\
	B_f & = & m_{f_3}\sqrt{\frac{r_{2}(r_{2}+r_{1}-1)(r_{2}+r_{2}-1)}{1-r_{2}},}\\
	C_f & = & m_{f_3}(r_{2}+r_{1}+r_{2}),\nonumber \\
	D_f & = & \sqrt{\frac{m_{f_1}m_{f_2}}{1-r_{2}}},\nonumber 
\end{eqnarray}
where $r_i=m_{f_i}/m_{f_3}$.
Applying a bi-unitary transformation, the fermion mass matrices are diagonalized. However, the Yukawa matrices generally remain non-diagonal:
\begin{equation}\label{eq:MfmatrixDiag}
	\overline{M}_f = V_{fL}^{\dagger} M_f V_{fR}= \tfrac{1}{\sqrt{2}} \left( v_1 \widetilde{Y}_{1}^{f} + v_2 \widetilde{Y}_{2}^{f} \right),
\end{equation}
where $\overline{M}_f = \operatorname{diag}(m_{f_1},m_{f_2},m_{f_3})$ and $\widetilde{Y}_{a}^{f} = V_{fL}^{\dag} Y_{a}^{f} V_{fR}$. As a consequence, flavor-changing neutral currents will be induced at tree-level.
The fact that $M_f$ is assumed as Hermitian, implies that $V_{fL}=V_{fR}\equiv V_f$. Here, $V_f=\mathcal{O}_{f}P_f$, where $P_f=\text{diag}\{e^{i\alpha_f},\,e^{i\beta_f},\,1\}$ and the $\mathcal{O}_{f}$ matrix given in Appx.~\ref{Omatrix}.

A fact of utmost importance is that $V_{u,d}$ must reproduce the measured matrix elements $V_{\text{CKM}}$, which can be obtained through $V_{\text{CKM}}=V_{u}^{\dagger}V_{d}$, as shown in Ref.~\cite{Arroyo-Urena:2013cyf}. From Eq.~\eqref{eq:MfmatrixDiag}, we obtain the following expression.
	
	\begin{equation}\label{RotateYukawas}
		\left[ \tilde{Y}_a^f \right]_{ij} = \frac{\sqrt{2}}{v_a}\delta_{ij}\bar{M}_{ij}^f-\frac{v_b}{v_a}\left[ \tilde{Y}_b^f \right]_{ij},
	\end{equation}
    where 
    \begin{equation}
     \left[ \tilde{Y}_b^f \right]_{ij}=\frac{\sqrt{m_im_j}}{v}  \left[\chi_b^f \right]_{ij}.
    \end{equation}
    In these expressions, the indices \(a\) and \(b\) label the two Higgs doublets and take the values \(1\) or \(2\), with \(a \neq b\). Different types of interactions (Type I, II, X, Y) can then be defined by specific assignments of the fermion couplings to \( \Phi_a \) or \( \Phi_b \) \cite{HernandezSanchez:2012eg}. The superscript $f$ stands for the flavor of fermions, while $\left[\chi_b^f \right]_{ij}$'s are unknown dimensionless parameters. In this paper, we select the following definitions
		\begin{eqnarray}\label{II}
			\left[ \tilde{Y}_1^d \right]_{ij} &=& \frac{\sqrt{2}}{v\cos\beta}\delta_{ij}\bar{M}_{ij}^d-\tan\beta\left[ \tilde{Y}_2^d \right]_{ij}\nonumber\\
			\left[ \tilde{Y}_2^u \right]_{ij} &=& \frac{\sqrt{2}}{v\sin\beta}\delta_{ij}\bar{M}_{ij}^u-\cot\beta\left[ \tilde{Y}_1^u \right]_{ij}\nonumber\\
			\left[ \tilde{Y}_1^\ell \right]_{ij} &=& \left[ \tilde{Y}_1^d \right]_{ij}(d\to\ell).
		\end{eqnarray}  
	From Eqs. \eqref{YukawaLagrangian}-\eqref{II}, we obtain the $\phi  \bar{f}_i f_j$ interactions:
	\begin{equation}
		\mathcal{L}_Y^\phi=\phi\bar{f}_i(S_{ij}^{\phi}+iP_{ij}^\phi\gamma^5)f_j,
	\end{equation}
	where $\phi=h,\,H,\,A$ and
	\begin{eqnarray}
		S_{ij}^{\phi}&=&\frac{gm_f}{2M_W}c_f^\phi\delta_{ij}+d^\phi_f\left[ \tilde{Y}_b^f \right]_{ij},\nonumber\\
		P_{ij}^\phi&=&\frac{gm_f}{2M_W}e_f^\phi\delta_{ij}+g^\phi_f\left[ \tilde{Y}_b^f \right]_{ij}.\label{SijPij}
	\end{eqnarray}
In the theoretical framework of the SM, $c_f^{\phi=h}=1$ and $d_f^{\phi=h}=e_f^{\phi=h}=g_f^{\phi=h}=0$. Meanwhile, for the 2HDM-III, these coefficients are presented in Table \ref{couplings}. 
		\begin{table}
        \begin{centering}
			\begin{tabular}{ccccccc}
				\hline 
				Coefficient & $c_{f}^{h}$ & $c_{f}^{A}$ & $c_{f}^{H}$ & $d_{f}^{h}$ & $d_{f}^{A}$ & $d_{f}^{H}$\tabularnewline
				\hline 
				\hline 
				$d$-type & $-\frac{\sin\alpha}{\cos\beta}$ & $-\tan\beta$ & $\frac{\cos\alpha}{\sin\beta}$ & $\frac{\cos(\alpha-\beta)}{\cos\beta}$ & $\csc\beta$ & $\frac{\sin(\alpha-\beta)}{\cos\beta}$\tabularnewline
				\hline 
				$u$-type & $\frac{\cos\alpha}{\sin\beta}$ & $-\cot\beta$ & $\frac{\sin\alpha}{\sin\beta}$ & $-\frac{\cos(\alpha-\beta)}{\sin\beta}$ & $\sec\beta$ & $\frac{\sin(\alpha-\beta)}{\sin\beta}$\tabularnewline
				\hline 
				leptons $\ell$ & $-\frac{\sin\alpha}{\cos\beta}$ & $-\tan\beta$ & $\frac{\cos\alpha}{\sin\beta}$ & $\frac{\cos(\alpha-\beta)}{\cos\beta}$ & $\csc\beta$ & $\frac{\sin(\alpha-\beta)}{\cos\beta}$\tabularnewline
				\hline 
			\end{tabular}
                            \par\end{centering}
			\caption{Coefficients for $\phi$-Fermion couplings in 2HDM-III with $CP$-conserving Higgs potential.}	\label{couplings}
			
		\end{table}

	
	\subsection{Constraints on the 2HDM-III parameter space}
    The parameters that have a direct impact on our predictions are the following:
	\begin{enumerate}
		\item Cosine of the mixing angles: $\cos(\alpha-\beta)$,
		\item Ratio of the VEV's: $\tan\beta$,
		\item The parameter responsible for the FCNI: $\chi_{sb}$. 
	\end{enumerate}
	
	A comprehensive analysis on the model parameter space is reported by one of us  in Ref.~\cite{Arroyo-Urena:2024soo}, including LHC Higgs boson data \cite{CMS:2017con, ATLAS:2019pmk} through so-called signals strengths. For a decay $S\to X$ and a production process $\sigma(pp\to S)$, the signal strength is parameterized as
	\begin{equation}\label{muX}
		\mathcal{\mu}_{X}=\frac{\sigma(pp\to h)\cdot\mathcal{BR}(h\to X)}{\sigma(pp\to h^{\text{SM}})\cdot\mathcal{BR}(h^{\text{SM}}\to X)},
	\end{equation}
	where $\sigma(pp\to S)$ is the production cross-section of $S$, with $S=h,\,h^{\text{SM}}$; here $h$ is the SM-like Higgs boson coming from an extension of the SM and $h^{\text{SM}}$ is the SM Higgs boson; $\mathcal{BR}(S\to X)$ is the branching ratio of the decay $S\to X$, with \( X = c\bar{c},\allowbreak\, b\bar{b},\allowbreak\, \tau^-\tau^+,\allowbreak\, \mu^-\mu^+,\allowbreak\, WW^*,\allowbreak\, ZZ^*,\allowbreak\, \gamma\gamma \). 
	We also confronted the free model parameters against upper limits on the branching ratio of lepton flavor-violating (LFV) processes, $\ell_i\to \ell_j\gamma$ \cite{Belle:2021ysv, BaBar:2009hkt, MEG:2016leq} and $\ell_i\to \ell_j\ell_k\bar{\ell}_k$~\cite{Workman:2022ynf}, and the decays $B_{(s,\,d)}^0\to\mu^+\mu^-$ \cite{CMS:2022mgd}\footnote{Analytical expressions for all the LFV processes and the decay $\mathcal{BR}(B_{(s,\,b)}\to \mu^-\mu^+)$ in the framework of 2HDM-III are also included in an investigation done by one of us~\cite{Arroyo-Urena:2024soo}.}.
	
\begin{figure}[!htb]
		\centering
		\includegraphics[width=8.5cm]{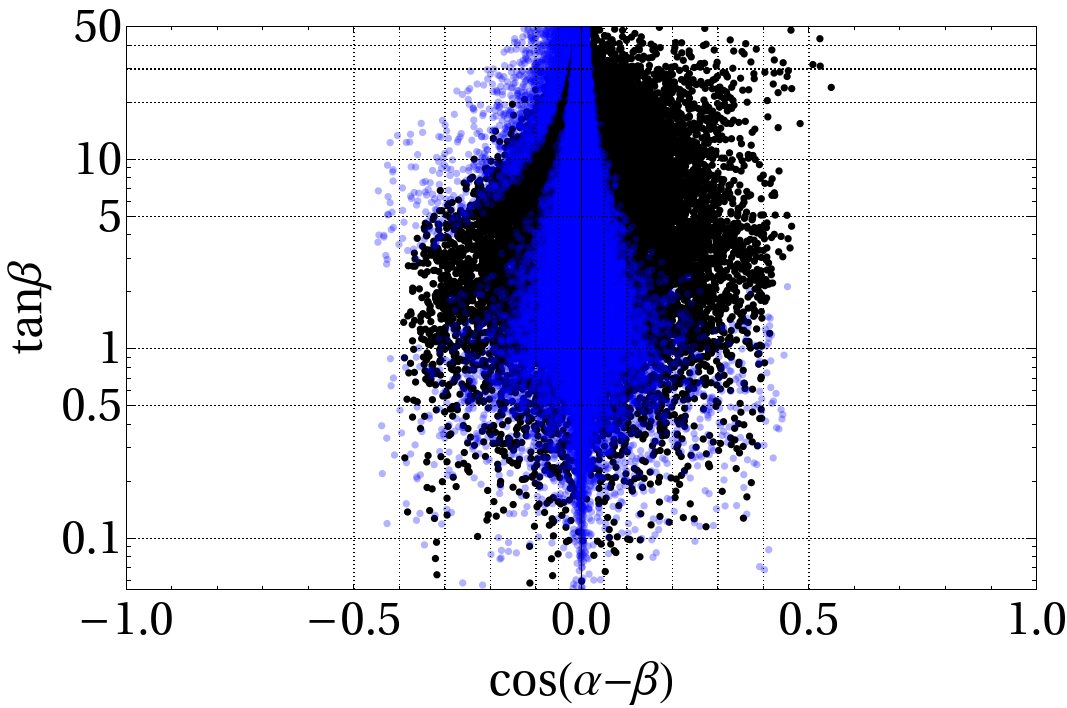}
		\caption{Allowed parameter space in the $\cos(\alpha-\beta)-\tan\beta$ plane. Blue (Black) points stand for those allowed values by LHC Higgs boson data (LFV processes).}\label{HixBosonData}
	\end{figure}

The parameter space allowed by the aforementioned observables, in the $\cos(\alpha-\beta)-\tan\beta$ plane, is presented in Fig.~\ref{HixBosonData}. The blue points represent those allowed by the signal strengths $\mu_X$ and the black points correspond to those allowed by the upper limits on branching ratios of the lepton flavor-violating processes.

\begin{figure}[!htb]
		\centering
		\includegraphics[width=7cm]{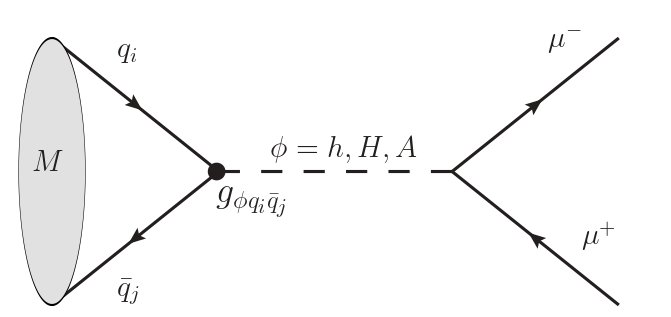}
		\caption{Generic Feynman diagram of the decay $M\to\mu^-\mu^+$, where $M=B_s$, $q_i=b$, and $q_j=s$. Such a decay is induced via $\phi=h,\,H,\,A$ in 2HDM-III.}\label{FDBmumu}
	\end{figure}

\begin{figure}[!htb]
	\centering
	\includegraphics[width=7cm]{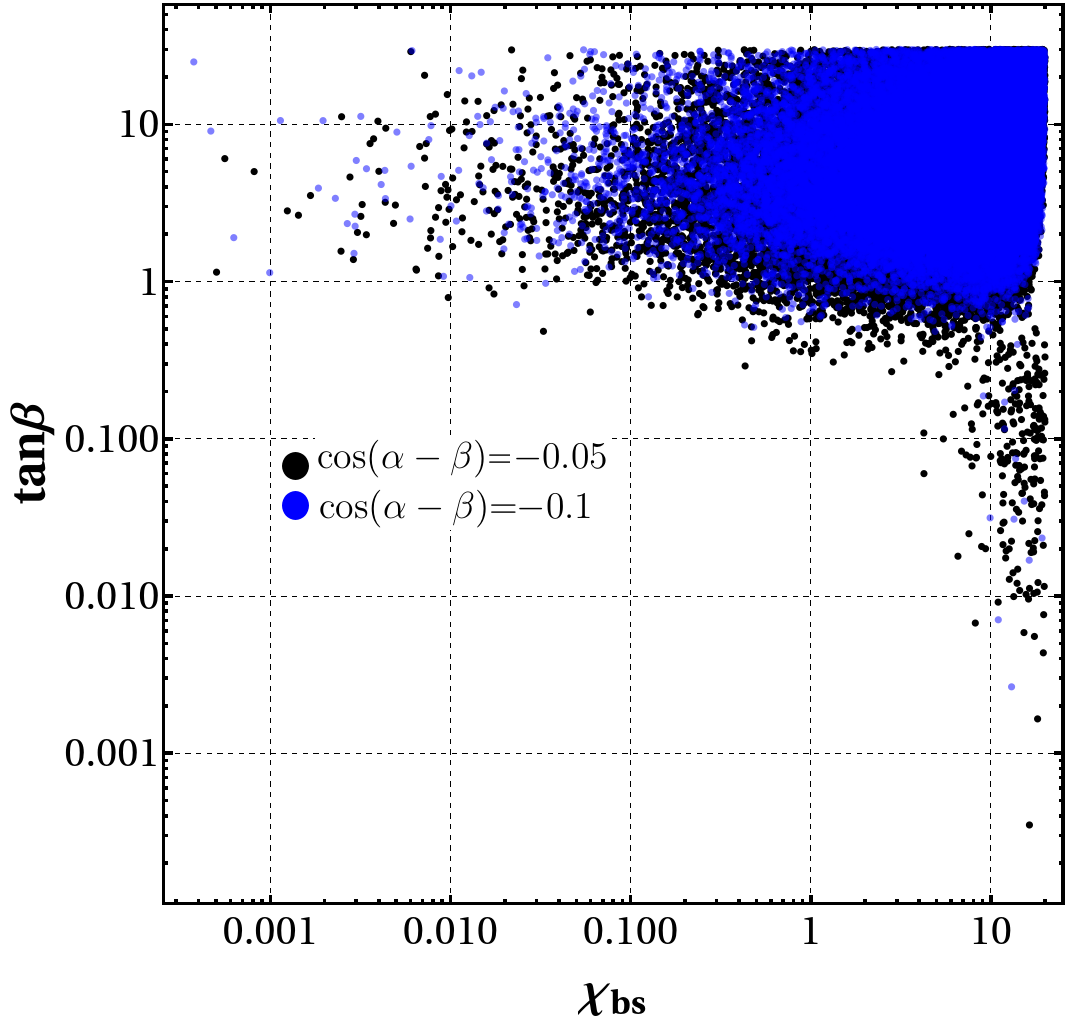}
	\caption{Allowed parameter space in the $\chi_{bs}-\tan\beta$ plane. Black ($\cos(\alpha-\beta)=-0.05$) and blue ($\cos(\alpha-\beta)=-0.1$) points stand for those allowed  values by experimental measurements on $\mathcal{BR}(B_{s}\to \mu^-\mu^+)$.}\label{chibs}
\end{figure}

It is of particular interest to highlight the process $\mathcal{BR}(B_s\to\mu^-\mu^+)$, as it can help us to derive constraints on the flavor-changing parameter $\chi_{bs}$. This parameter ---in addition to $\tan\beta$ and $\cos(\alpha-\beta)$--- directly affects the $\phi q_iq_j$ interaction, which is illustrated in Fig. \ref{FDBmumu}.  Therefore, we present in Fig.~\ref{chibs} the $\chi_{bs}-\tan\beta$ plane, where the black and blue points are those allowed by the experimental measurement of $\mathcal{BR}(B_{s}\to \mu^-\mu^+)$, and correspond to $\cos(\alpha-\beta)=-0.05$ and $\cos(\alpha-\beta)=-0.1$, respectively.

The parameter space reported was identified by performing a systematic scan over the ranges outlined in Table \ref{scan_parameters}. 

\begin{table}

\caption{Parameter scan for those model parameters directly influencing the LFV processes (left) and the signal strength $\mu_{X}$ (right). A total of $10^{8}$
random points were generated for each case.}\label{scan_parameters}

\begin{centering}
\begin{tabular}{|c|c|}
\hline 
Parameter & Scanned range\tabularnewline
\hline 
\hline 
$\cos(\alpha-\beta)$ & $[-1,\,1]$\tabularnewline
\hline 
$\tan\beta$ & $[0.1,\,50]$\tabularnewline
\hline 
$\chi_{tt},\,\chi_{\tau\mu},\,\chi_{\tau e},\,\chi_{\mu e},\,\chi_{\tau\tau},\,\chi_{\mu\mu}$ & $[-1,\,1]$\tabularnewline
\hline 
$M_{A},\,M_{H},\,M_{H^{\pm}}$ & $[500,\,1000]$ (GeV)\tabularnewline
\hline 
\end{tabular} %
\begin{tabular}{|c|c|}
\hline 
Parameter & Scanned range\tabularnewline
\hline 
\hline 
$\cos(\alpha-\beta)$ & $[-1,\,1]$\tabularnewline
\hline 
$\tan\beta$ & $[0.1,\,50]$\tabularnewline
\hline 
$\chi_{tt},\,\chi_{bb},\,\chi_{\tau\tau}$ & $[-1,\,1]$\tabularnewline
\hline 
$M_{H^{\pm}}$ & $[500,\,1000]$ (GeV)\tabularnewline
\hline 
\end{tabular}
\par\end{centering}
\end{table}

The parameters $\chi_{\tau\mu}$, $\chi_{\tau e}$, $\chi_{\mu e}$, $\chi_{\tau\tau}$, and $\chi_{\mu\mu}$ directly influence decays induced by the interactions $\phi \tau\mu$, $\phi \tau e$, $\phi \mu e$, $\phi\tau\tau$, and $\phi \mu\mu$. In particular, all of them generate the decays $\ell_i \to \ell_j \gamma$ at one-loop level, whereas $\phi tt$ contributes to the same processes but at two-loop order through Barr–Zee diagrams. The masses of the (pseudo)scalar bosons $M_A$, $M_H$, and $M_{H^{\pm}}$ also play a role in these decays, as well as in the process $B_s \to \mu^-\mu^+$. Nevertheless, within the scanned parameter space, the dominant contribution arises from the Higgs boson.

Once we show viable regions of the free model parameters, we now turn to present analytical expressions for the branching ratio $\mathcal{BR}(h\to bs)$ in terms of the parameters $\chi_{bs}$, $\tan\beta$ and $\cos(\alpha-\beta)$. The partial decay width $\Gamma(h\to bs)$ reads:
\begin{equation}
 \Gamma(h\to bs) = \frac{3m_h}{128\pi}\, g_{hbs}^2 \,
    \Big[4-\big(\sqrt{\tau_b}+\sqrt{\tau_s}\big)^{2}\Big]^{3/2}\Big[4-\big(\sqrt{\tau_b}-\sqrt{\tau_s}\big)^{2}\Big]^{1/2},    
\end{equation}
where the coupling $g_{hbs}$ is given by
\begin{equation}\label{hbs_coupling}
    g_{hbs} = \frac{\cos(\alpha-\beta)\tan\beta}{\sqrt{2}\sin\beta} \,
    \frac{\sqrt{m_s m_b}}{v} \, \chi_{bs},
\end{equation}
and $\tau_{(b,s)} = 4 m_{(b,s)}^2 / m_h^2$. The corresponding branching ratio is
\begin{equation}
    \mathcal{BR}(h\to bs) = \frac{\Gamma(h\to bs)}{\underset{i}{\sum}{\Gamma_i}},
\end{equation}
where $\underset{i}{\sum}{\Gamma_i}$ is the total Higgs width decay.
	

	\section{Collider analysis}\label{SecIII}
   In this section, we present the strategy for separating the signal from the background processes. We first use a Monte Carlo generator to obtain a sample of simulated events. The simulations of the phenomenological processes were performed by using $\texttt{FeynRules}$~\cite{Alloul:2013bka} to build the 2HDM-III and
	produce the UFO files~\cite{Degrande:2011ua}, subsequently we generated parton-level events for both the signal and the SM background processes using $\texttt{MadGraph5}$ \cite{MadGraphNLO}, later we performed shower and hadronization with \texttt{Pythia8} \cite{Sjostrand:2008vc}, and finally the detector response was emulated using \texttt{Delphes 3} \cite{delphes}. 
    According to the information contained in the \texttt{Delphes} cards\footnote{We use the default \texttt{Delphes} cards for CEPC and FCC-ee,  $\texttt{delphes\_card\_CEPC.tcl}$ and $\texttt{delphes\_card\_IDEA.tcl}$, respectively.}, the $b-$tagging efficiency is $\epsilon_b=0.8$, the probability that a $c-$jet or any other light jet $j$ is mistagged as a $b-$jet is 0.1 and 0.001, respectively. 
	\subsection{Signal and background}
	
	\begin{itemize}
		\item \textit{Signal:} We search in the final state a pair of jets (one of them tagged as a $b-$jet) and a pair of charged leptons ($\ell^-\ell^+$), which are product of the decays $h\to bs$ ($bs=b\bar{s}+\bar{b}s$) and $Z\to\ell^-\ell^+$ $(\ell=e,\mu)$, respectively. The Feynman diagram of the signal is presented in Fig.~\ref{FDsignal}.
        \begin{figure}[!htb]
		\centering
		\includegraphics[width=6cm]{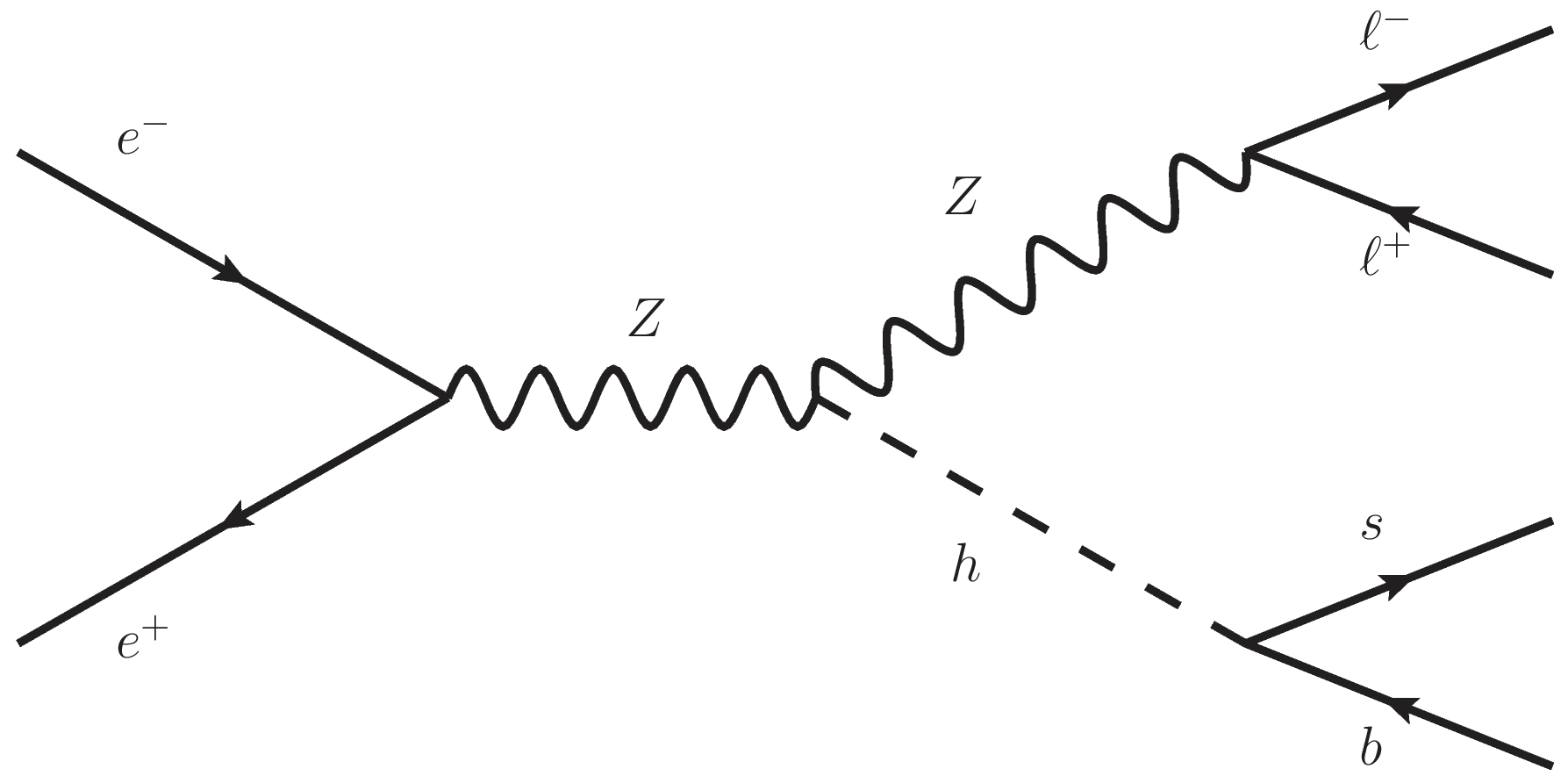}
		\caption{Feynman diagram of the signal.}\label{FDsignal}
	\end{figure}
    \\
We observe from Fig.~\ref{FDsignal} that there are two non standard interactions,  $hbs$ and $hZZ$, which are given by the following couplings:
\begin{eqnarray}
     g_{hbs}&=&\frac{\cos(\alpha-\beta)\tan\beta}{\sqrt{2}\sin\beta}\frac{\sqrt{m_s m_b}}{v}\chi_{bs},\\
     g_{hZZ}&=&\sin(\alpha-\beta)\frac{g}{\cos\theta_W}m_Z,
\end{eqnarray}\label{Coupling_hbs}
where $\theta_W$ is the weak mixing angle. From the coupling $g_{hbs}$ we note that the signal increases for high values of $\tan\beta$ and $\chi_{bs}$. In the following analysis we will exploit this feature to maximize the signal significance, but maintaining consistency with the findings reported in the previous section.

Figure~\ref{XS_signal} presents the production cross section of the signal as a function of the flavor-changing parameter $\chi_{bs}$ and $\tan\beta$, for (a) $\cos(\alpha-\beta)=-0.05$ at $\sqrt{s}=240$ GeV, (b) $\cos(\alpha-\beta)=-0.1$ at $\sqrt{s}=240$ GeV, (c) $\cos(\alpha-\beta)=-0.05$ at $\sqrt{s}=365$ GeV, and (d) $\cos(\alpha-\beta)=-0.1$ at $\sqrt{s}=365$ GeV. The specific values $\cos(\alpha-\beta) = -0.05,\,-0.1$ not only satisfy the experimental constraints discussed in the previous section but are also motivated by the decoupling limit (along with the condition $M_\phi\gg v$), wherein the Standard Model emerges as the limiting case. It is worth noting that, in principle, the ranges $\chi_{bs}\in(0,10]$ and $\tan\beta\in(0,20]$ might be expected to yield large —and potentially problematic— values of the coupling $g_{hbs}$, which could violate the perturbativity bound $g_{hbs}\leq 4\pi$. However, a quick numerical assessment indicates the opposite: for instance, taking $\chi_{bs}=10$, $\tan\beta=20$, and $\cos(\alpha-\beta)=-0.1$, one finds $|g_{hbs}|=0.036$, well below that limit.
    \begin{figure}[!htb]
		\centering
		\subfigure[]{\includegraphics[width=7cm]{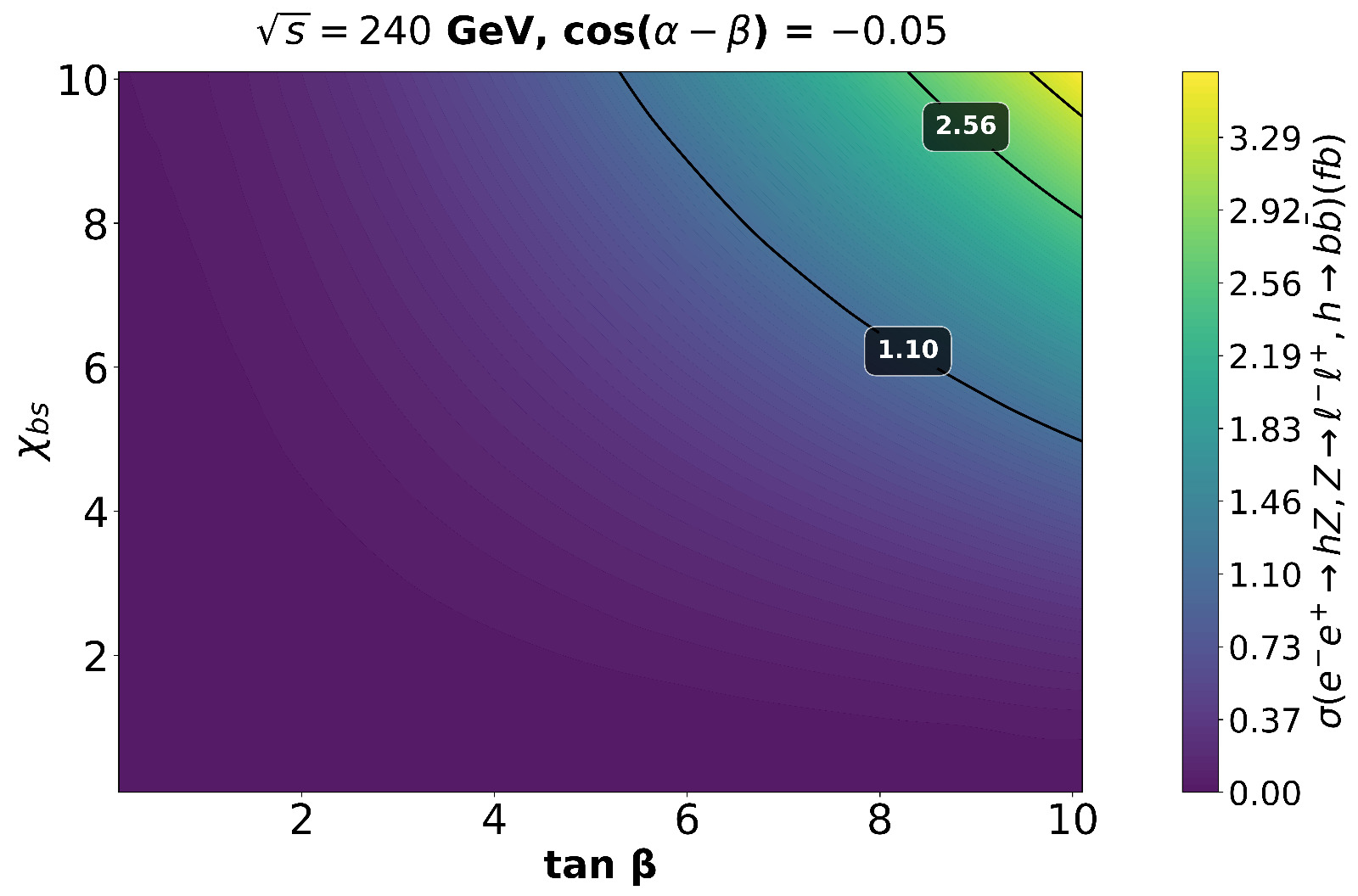}}
        \subfigure[]{\includegraphics[width=7cm]{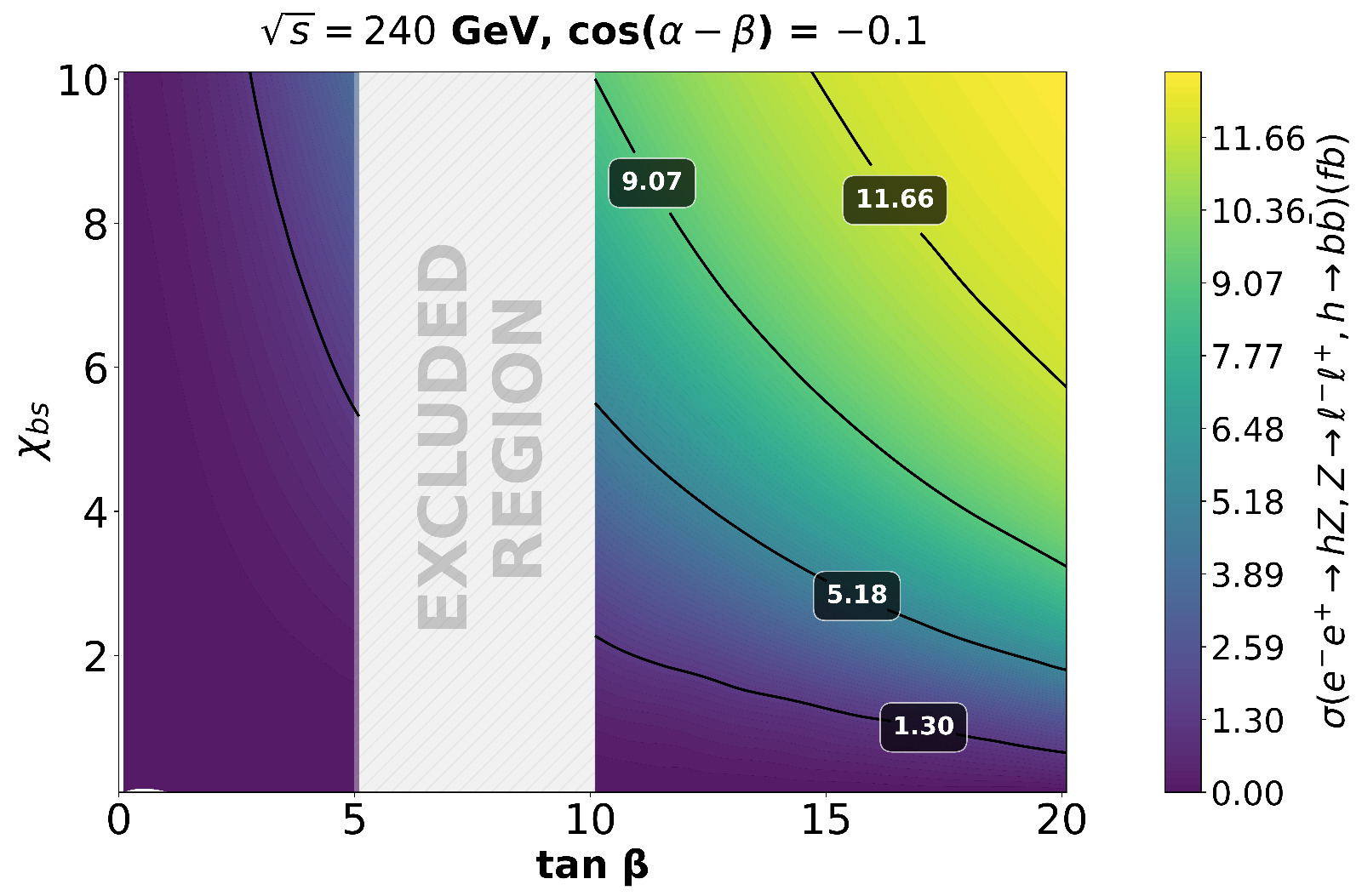}}
        \subfigure[]{\includegraphics[width=7cm]{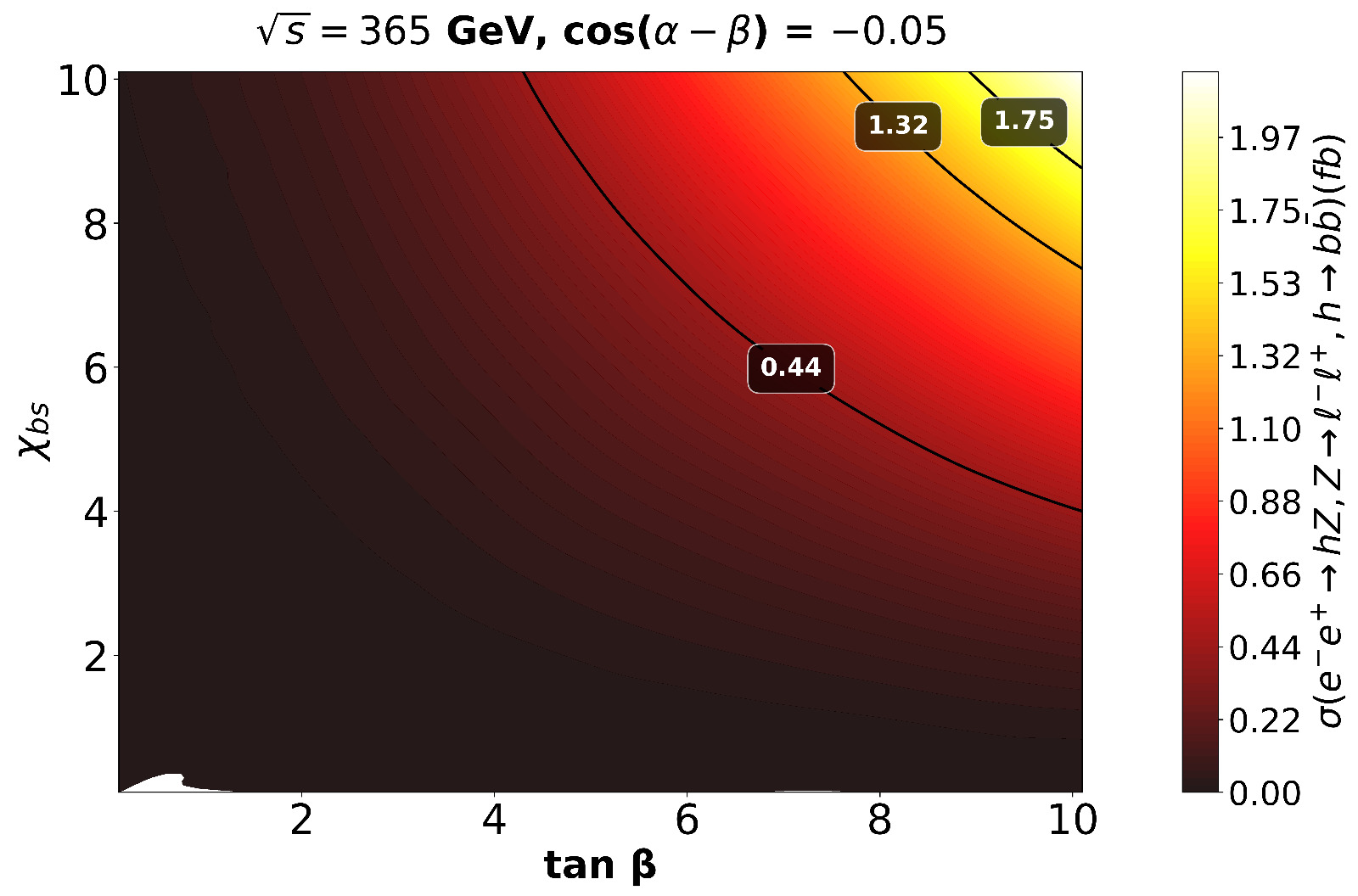}}
        \subfigure[]{\includegraphics[width=7cm]{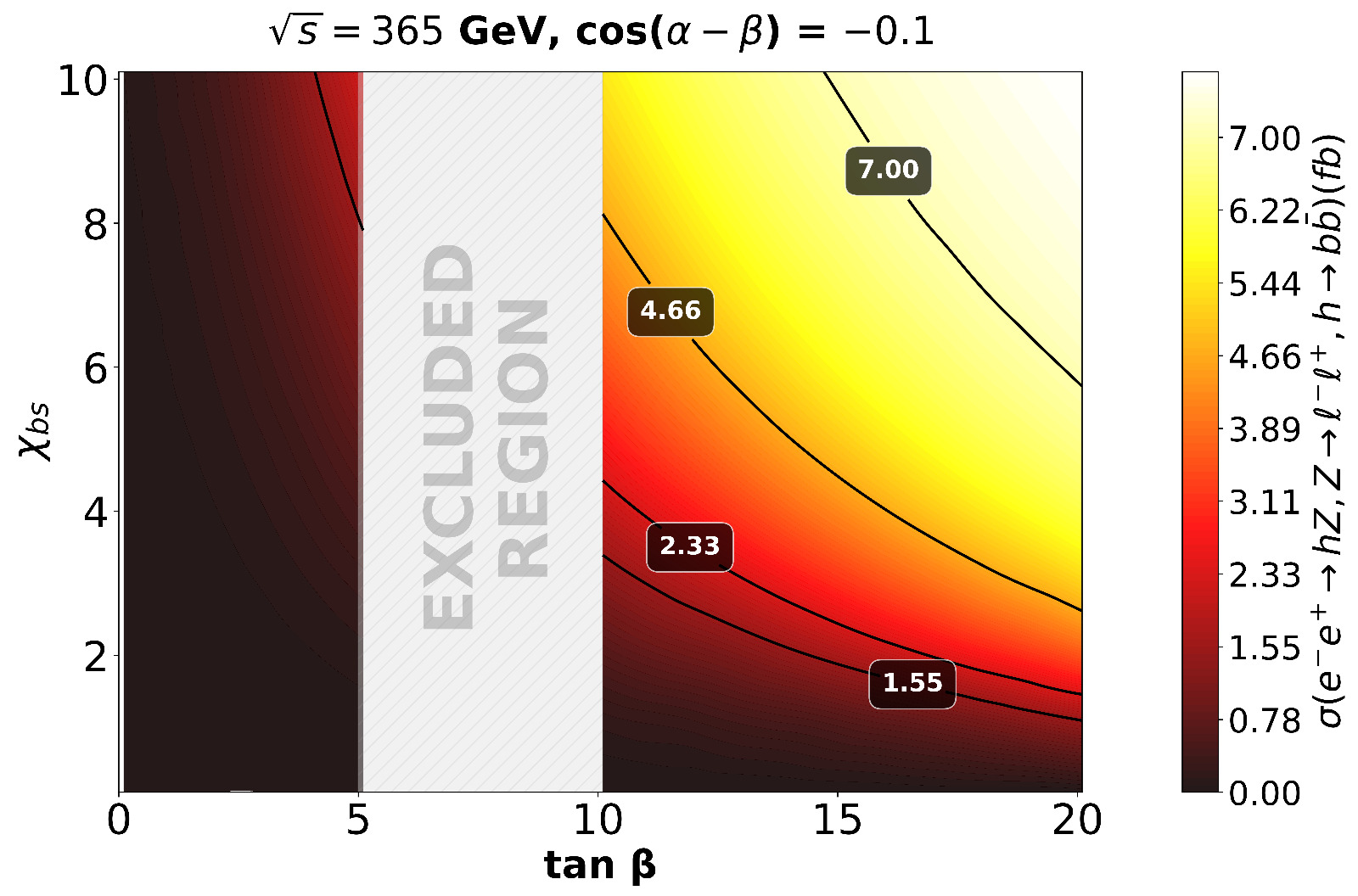}}
		\caption{Cross section of the signal as a function of flavor-changing parameter $\chi_{bs}$ and $\tan\beta$.}\label{XS_signal}
	\end{figure}
   

		\item \textit{Background:} The SM background processes that hide the signal come mainly from:
        \begin{itemize}
            \item $e^-e^+\to Zh$, with subsequent decays:
            \begin{itemize}
                \item  $Z\to\ell^-\ell^+,\,h\to jj$.
            \end{itemize}
            \item $e^-e^+\to ZZ$, 
            \begin{itemize}
                \item $Z \to \ell^-\ell^+,\,Z\to jj$.
            \end{itemize}
            \item $e^-e^+\to e^-e^+Z$, 
            \begin{itemize}
                \item $Z\to jj$.
            \end{itemize}
            \item $e^-e^+\to t\bar{t}\to W^+jW^-j\to\ell^-\bar{\nu}_\ell j \ell^+\nu_{\ell}j$,
        \end{itemize}
where $j=u,\,\bar{u},\,d,\,\bar{d},\,c,\,\bar{c},\,s,\,\bar{s},\,b,\,\bar{b}$ and $\ell^{\pm}=e^{\pm},\,\mu^{\pm}$. It should be noted that the last background process is relevant only for a center-of-mass energy $\sqrt{s}\geq 2m_{\rm top}$, so for the case of $\sqrt{s}=240$ GeV it will not be important.  Numerical cross sections of the SM background processes at $\sqrt{s}=240$ GeV and at $\sqrt{s}=365$ GeV are given in Tables~\ref{tb:XS-BGD240} and \ref{tb:XS-BGD350}, respectively.
\begin{table}[!htb]
\caption{Cross sections of the dominant SM background.}\label{tb:XS-BGD240}
\begin{centering}
\begin{tabular}{|c|c|}
\hline 
Cross-section & $\sqrt{s}=240$ GeV\tabularnewline
\hline 
\hline
$\sigma(e^{-}e^{+}\to Zh \to \ell\bar{\ell}jj)$ & 13.15 fb\tabularnewline
\hline
$\sigma(e^{-}e^{+}\to ZZ\to \ell\bar{\ell}jj)$ & 106.29 fb \tabularnewline
\hline 
$\sigma(e^{-}e^{+}\to e^{-}e^{+}Z\to e^{-}e^{+}jj)$ & 105.74 fb \tabularnewline
\hline 
\end{tabular}
\par\end{centering}
\end{table}
\begin{table}[!htb]
	\caption{Cross sections of the dominant SM background.}\label{tb:XS-BGD350}
	\begin{centering}
		\begin{tabular}{|c|c|}
			\hline 
			Cross-section & $\sqrt{s}=365$ GeV\tabularnewline
			\hline 
			\hline
			$\sigma(e^{-}e^{+}\to Zh \to \ell\bar{\ell}jj)$ & 6.527 fb\tabularnewline
			\hline
			$\sigma(e^{-}e^{+}\to ZZ\to \ell\bar{\ell}jj)$ & 59.01 fb \tabularnewline
			\hline 
			$\sigma(e^{-}e^{+}\to e^{-}e^{+}Z\to e^{-}e^{+}jj)$ & 92.93 fb \tabularnewline
			\hline 
			$\sigma(e^{-}e^{+}\to t \bar{t}\to \ell\bar{\ell}jj \nu_\ell\bar{\nu_\ell})$ & 10.77 fb\tabularnewline
			\hline
		\end{tabular}
		\par\end{centering}
\end{table}
 Note that the cross section of the first background process ($e^{-}e^{+}\to Zh$) is an order of magnitude smaller than all the others. However, it is an irreducible process that is the main source of signal contamination. 
	
\subsection{Signal–background discrimination}    
While baseline kinematic selections are applied (e.g. $30\,\text{GeV}<p_T^{b-\rm{jet}},\,p_T^{j}$ , $20\,\text{GeV}<p_T^{\ell^-},\,p_T^{\ell^+}$), the primary initial background rejection is achieved through two resonant mass windows: $80<M_{\rm inv}(\ell^-\ell^+)<100$ GeV and $105<M_{\rm inv}(bs)<145$ GeV. These directly target the signature $Z$ and Higgs boson resonances characteristic of the signal. Nevertheless, these cuts, while selective, do not yield a sufficiently pure sample. 
The signal-to-background ratio remains low due to irreducible and sophisticated backgrounds that survive the mass constraints. Therefore, to achieve the necessary discrimination power, we employ a Boosted Decision Tree (BDT)~\cite{Coadou:2022nsh, Hastie:2009itz} trained on a broader set of kinematic observables, which is presented as the final and most effective step of our analysis.
\end{itemize}
    \subsubsection{Multivariate Analysis}
\label{sec:mva}

\begin{figure}[htbp]
\centering
\subfigure[]{\includegraphics[width=0.48\textwidth]{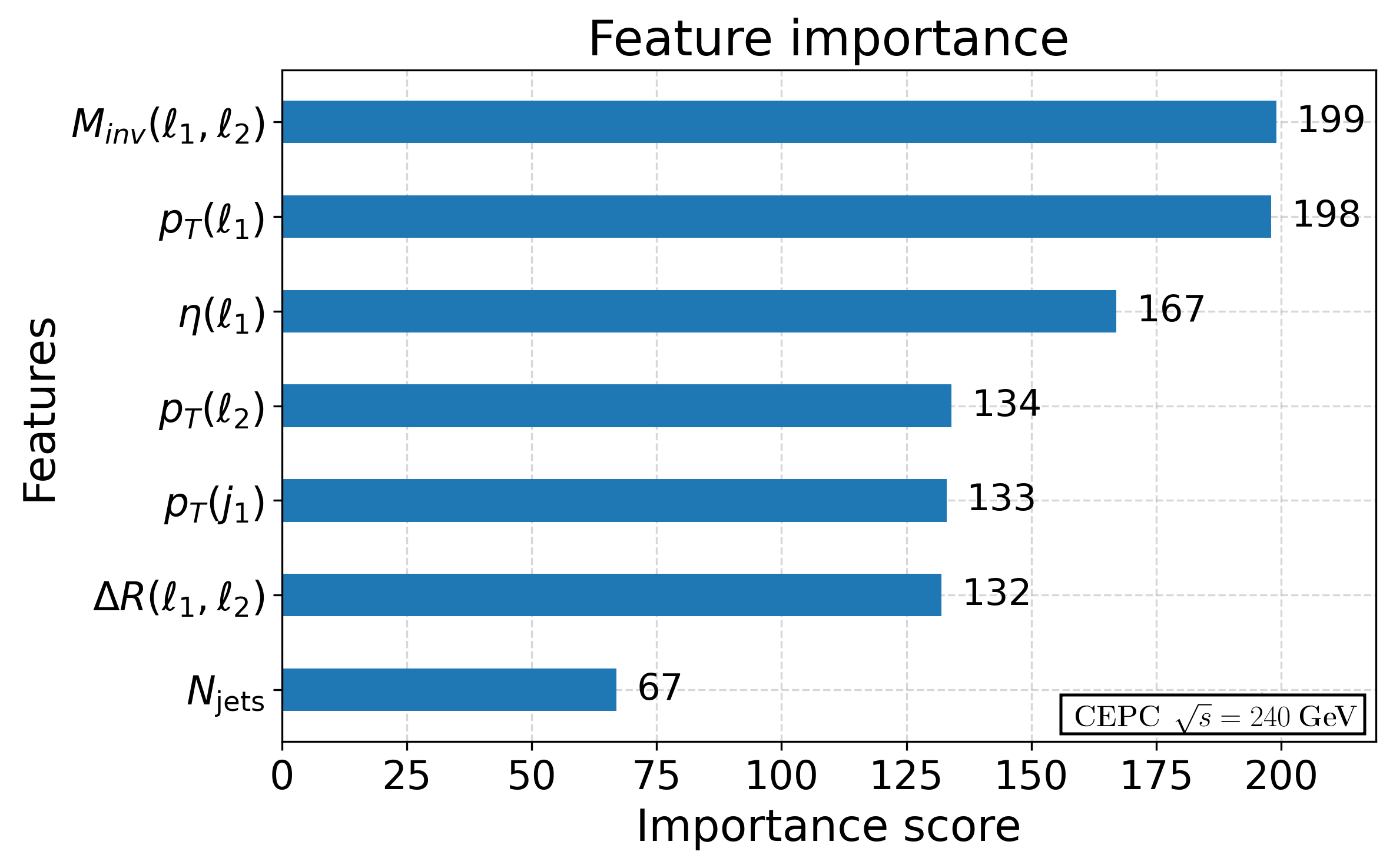}}
\subfigure[]{\includegraphics[width=0.48\textwidth]{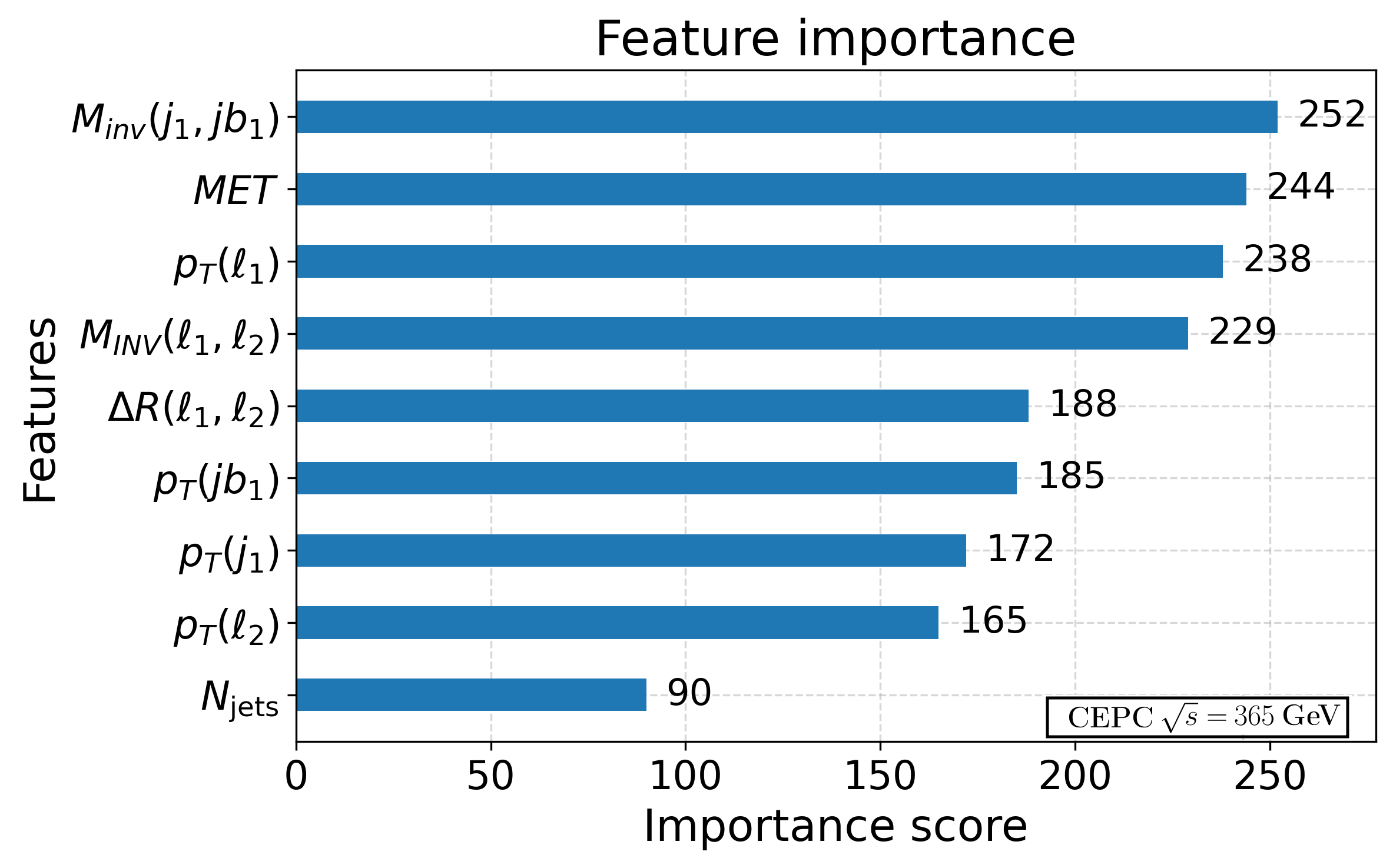}}
\subfigure[]{\includegraphics[width=0.48\textwidth]{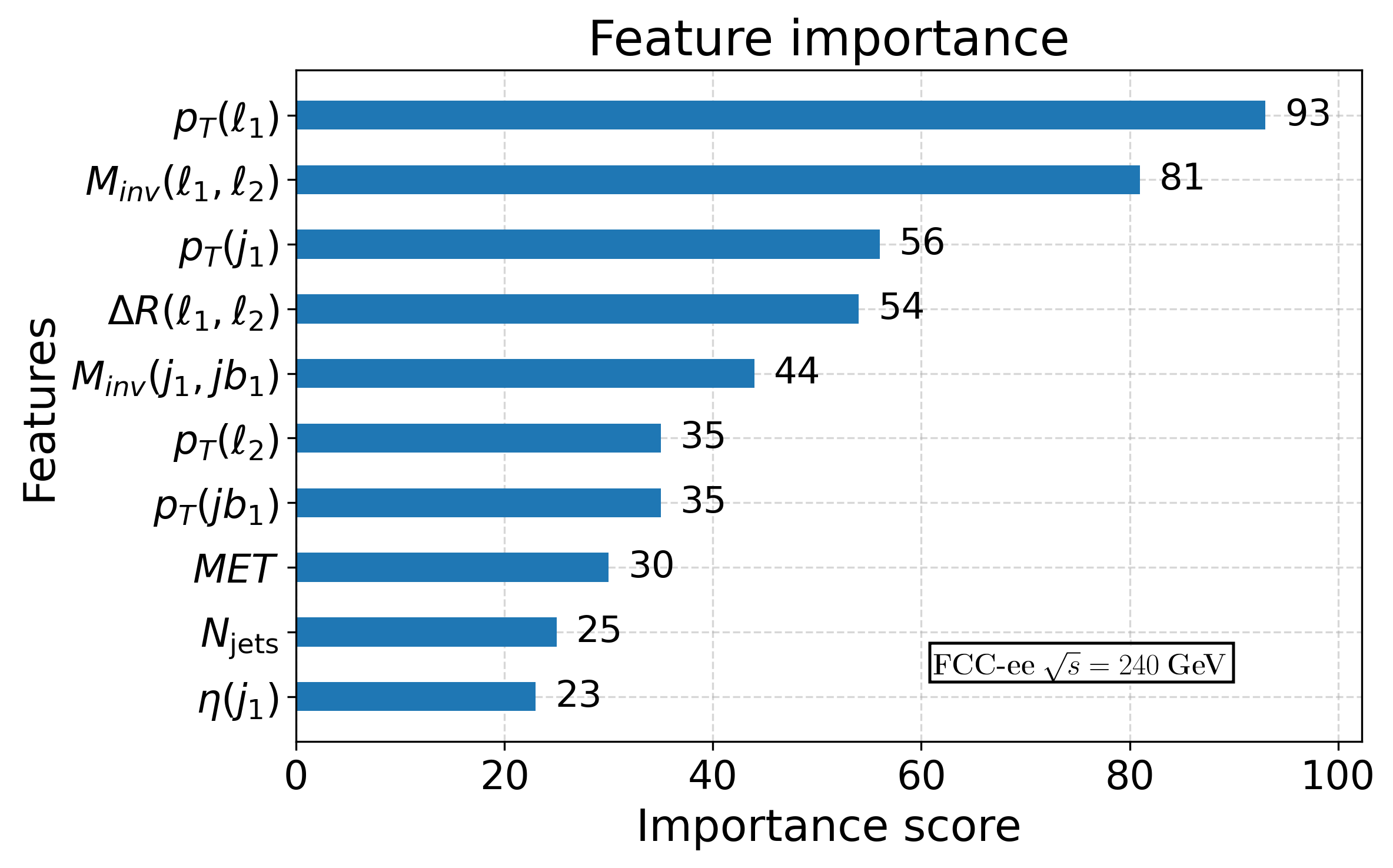}}
\subfigure[]{\includegraphics[width=0.48\textwidth]{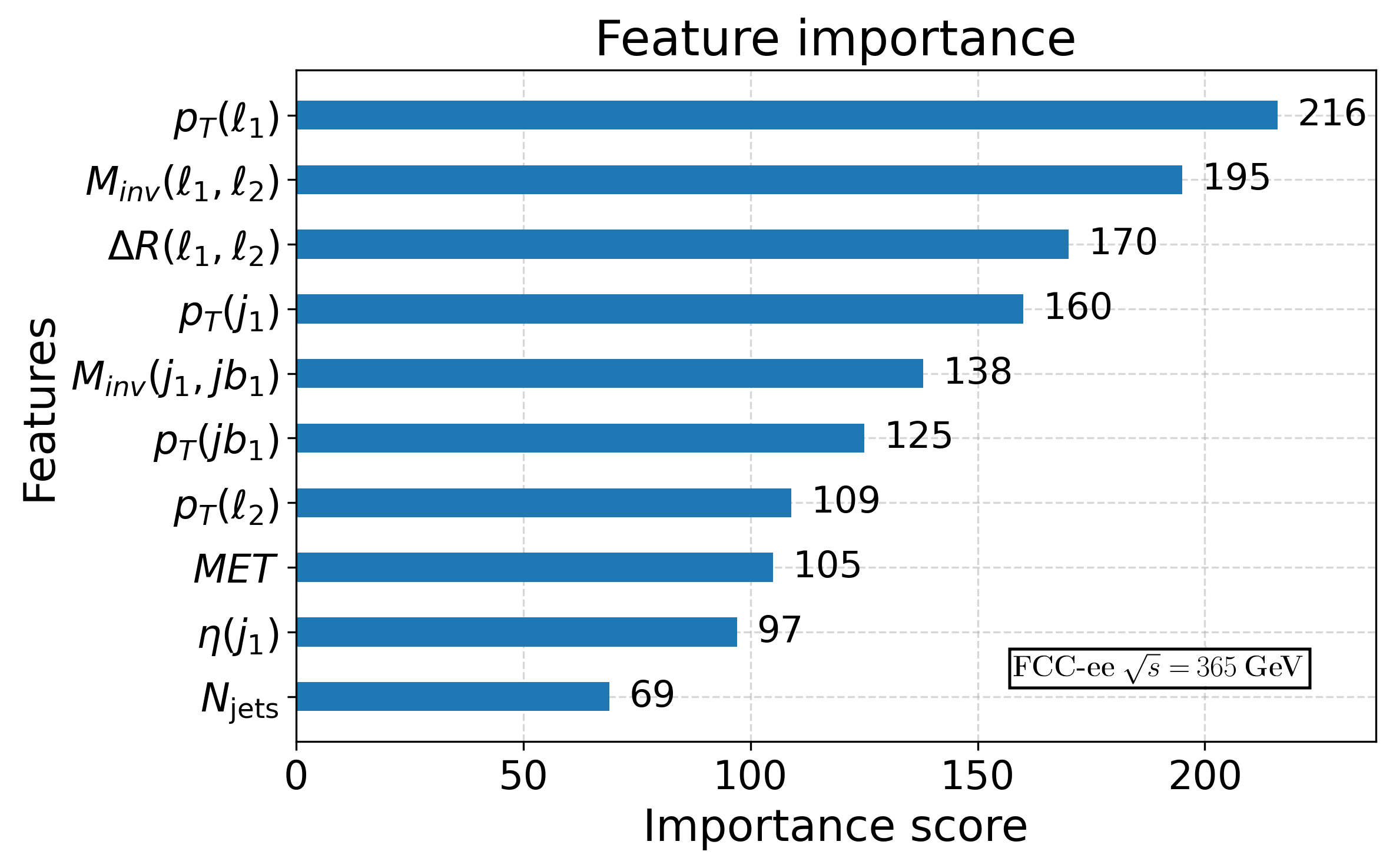}}
\caption{Ranking of the observables used for the BDT analysis. The index 1 stands for that with larger transverse momentum, while the index 2 is assigned to the sub-leading particle.}
\label{features}
\end{figure}

We implemented a selection based on Multivariate Analysis (MVA) discriminators, which combine several observables into a single, a more powerful classifier. For the MVA training, we employed the BDT algorithm~\cite{Book:bdt}, utilizing the XGBoost library~\cite{Chen:2016:XST:2939672.2939785} and its advanced gradient boosting framework. The BDT classifiers were trained on the kinematic observables of the final-state particles, as shown in Table~\ref{tab:Variables}. Each scenario was analyzed separately to achieve good performance. The feature importance for the four cases is illustrated in Fig.~\ref{features}.

\begin{table}
\caption{List of the variables used to train and test the signal and background events.}\label{tab:Variables}
\centering
\begin{tabular}{@{} l p{0.6\linewidth} @{}}
\toprule
Variable & Description \\
\midrule
$M_{\text{inv}}(\ell^-, \ell^+)$ & Invariant mass of the leptons $\ell^-=e^-,\,\mu^-$ and $\ell^+=e^+,\,\mu^+$ decaying from the $Z$ boson. \\
$M_{\text{inv}}(j_{1}, jb_{1})$ & Invariant mass of a light jet $j_{1}$ and a b-jet $jb_{1}$ decaying from the Higgs boson. \\
$p_{T}(p_{i})$ & Transverse momentum of $p_{i}=\ell^-,\ell^+,j_{1},jb_{1}$. \\
$\eta(p_{i})$ & Pseudorapidity of $p_{i}=\ell^-,\ell^+,j_{1},jb_{1}$. \\
$\Delta R(\ell^-, \ell^+)$ & Angular separation between $\ell^-$ and $\ell^+$. \\
$N_{\text{jets}}$ & Number of jets per event. \\
MET & Missing transverse energy. \\
\bottomrule
\end{tabular}
\end{table}

Training was performed using Monte Carlo simulated samples, which include the dominant irreducible backgrounds. All signal and background samples were generated with $5\times 10^5$ events each, ensuring a balanced training set and robust statistical validation.

\begin{figure}[htbp]
\centering
\subfigure[]{\includegraphics[width=0.45\textwidth]{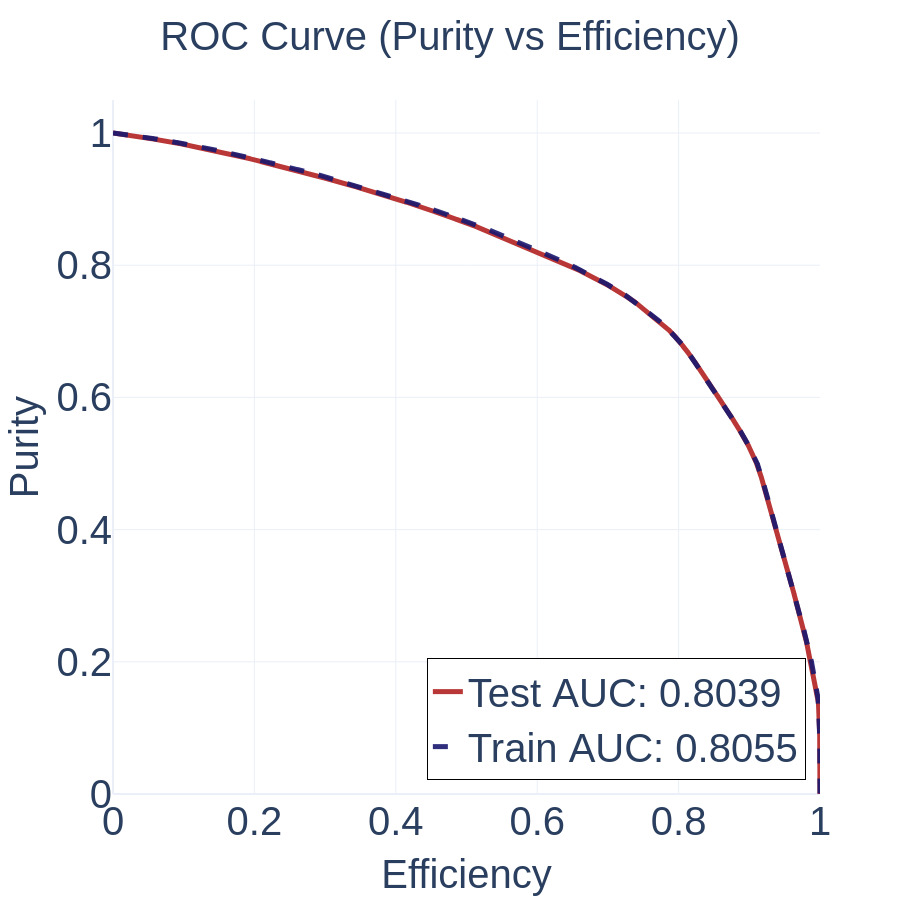}}
\subfigure[]{\includegraphics[width=0.45\textwidth]{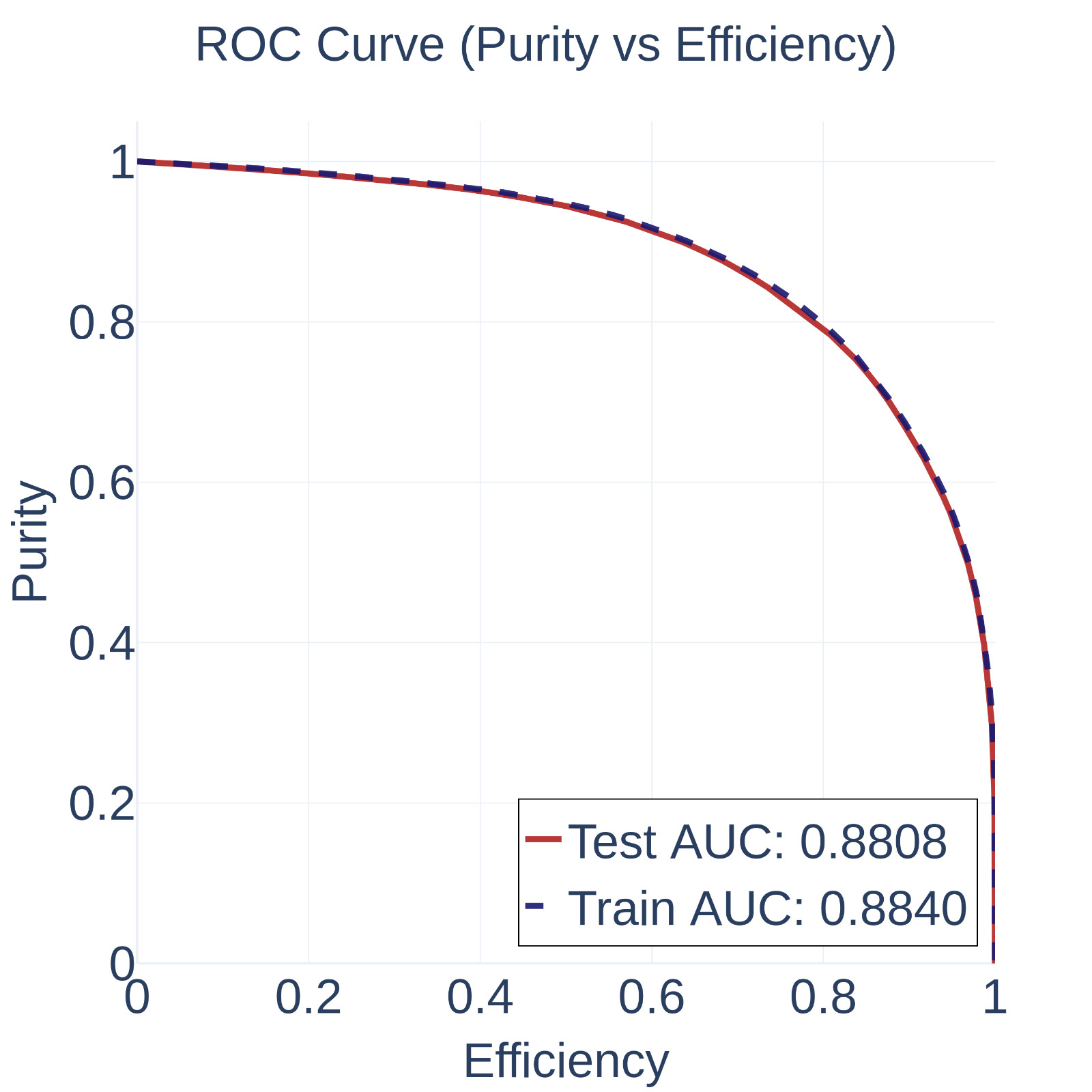}}
\subfigure[]{\includegraphics[width=0.45\textwidth]{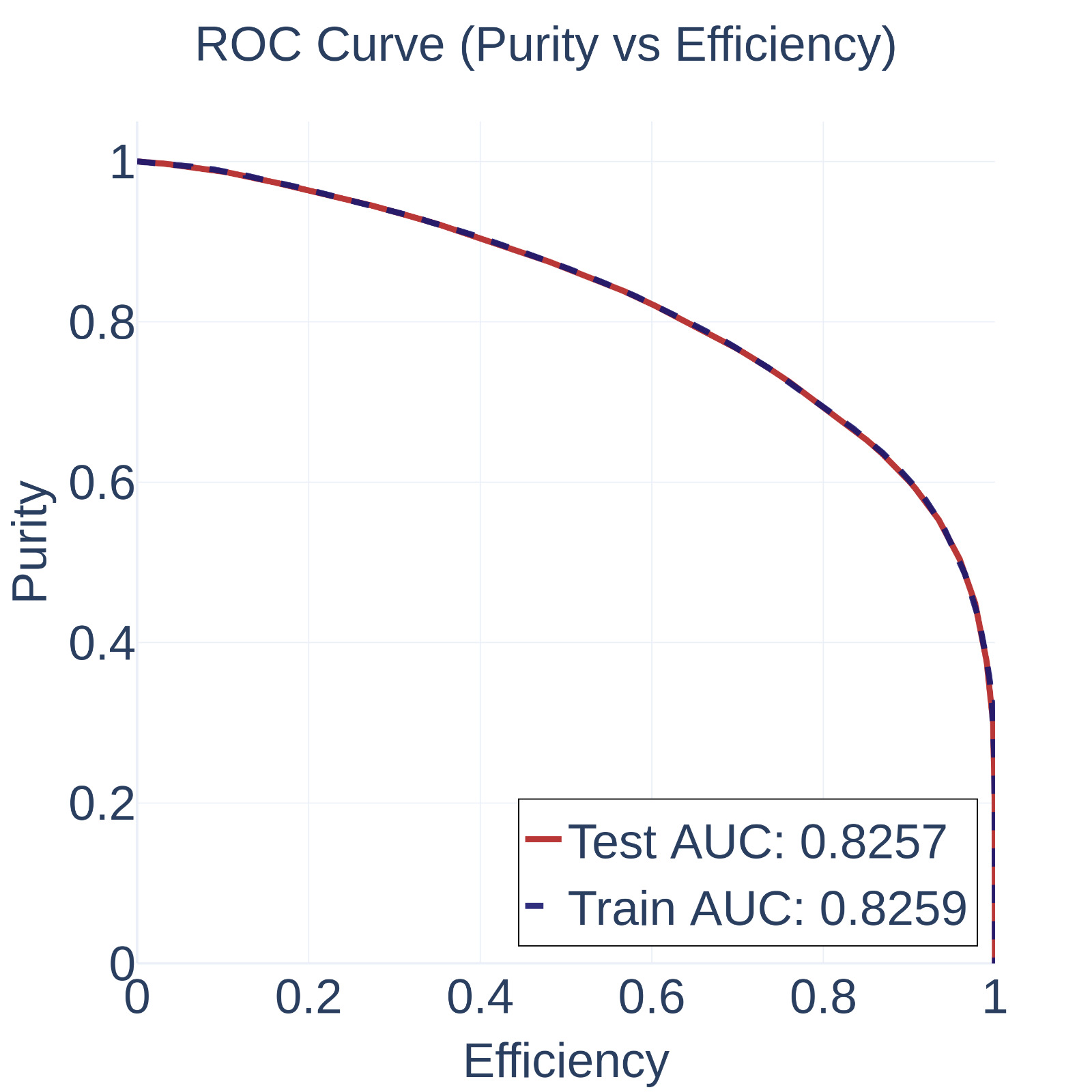}}
\subfigure[]{\includegraphics[width=0.45\textwidth]{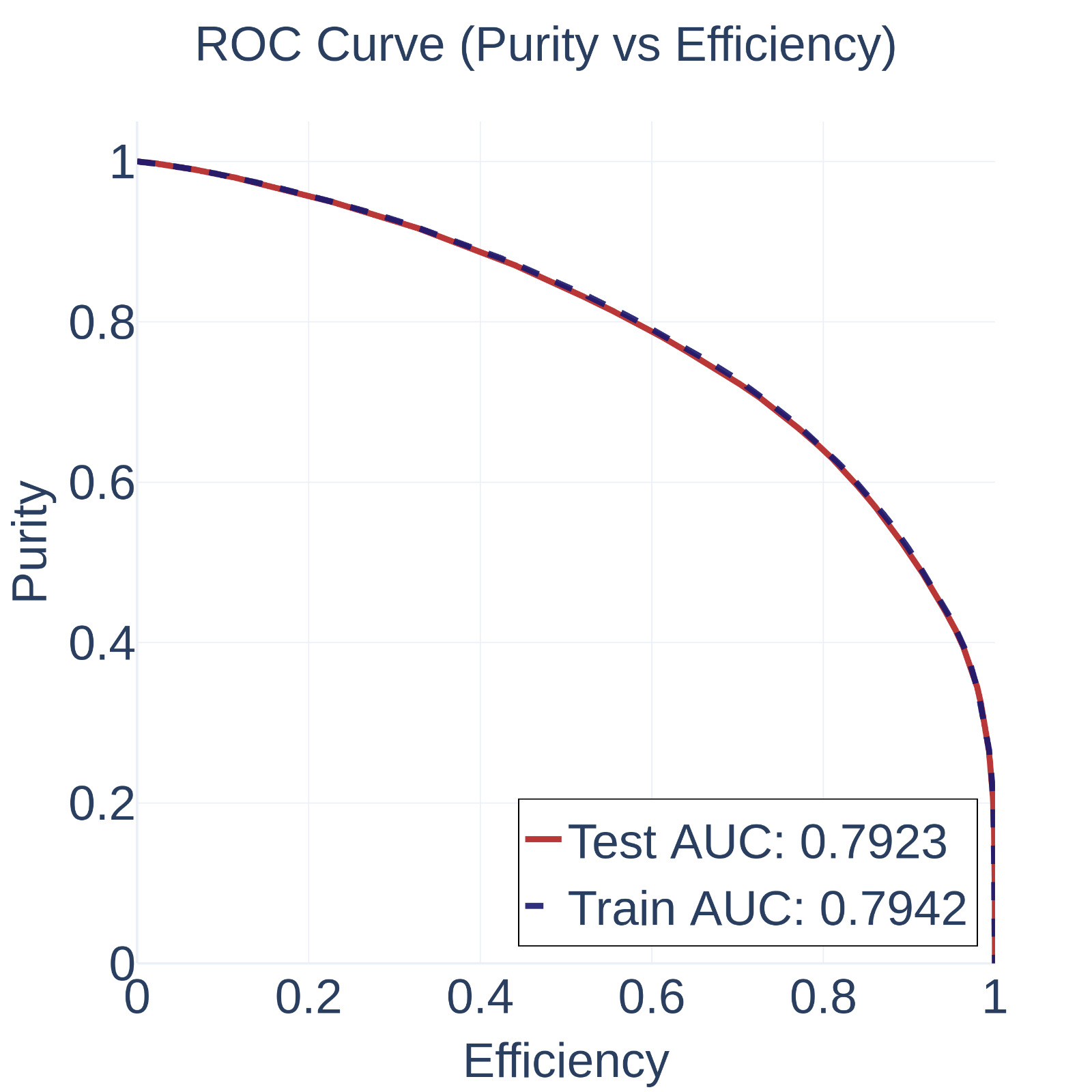}}
\caption{ROC curves (Purity vs. Efficiency) for the trained BDT classifier. (a) CEPC at $\sqrt{s}=240$ GeV, (b) CEPC at $\sqrt{s}=365$ GeV, (c) FCC-ee at $\sqrt{s}=240$ GeV, and (d) FCC-ee at $\sqrt{s}=365$ GeV }
\label{fig:roc}
\end{figure}

\begin{figure}[htbp]
\centering
\subfigure[]{\includegraphics[width=0.495\textwidth]{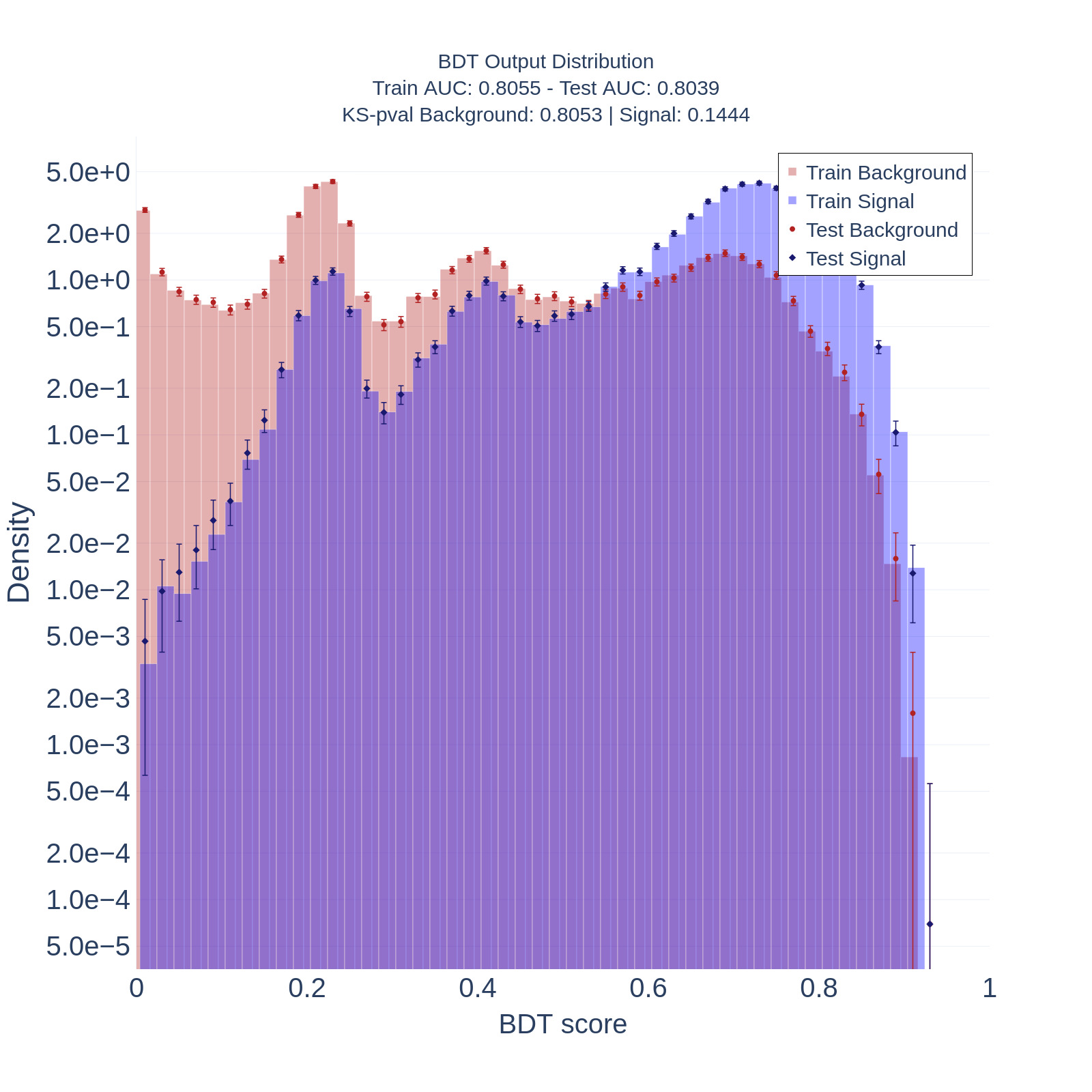}}
\subfigure[]{\includegraphics[width=0.495\textwidth]{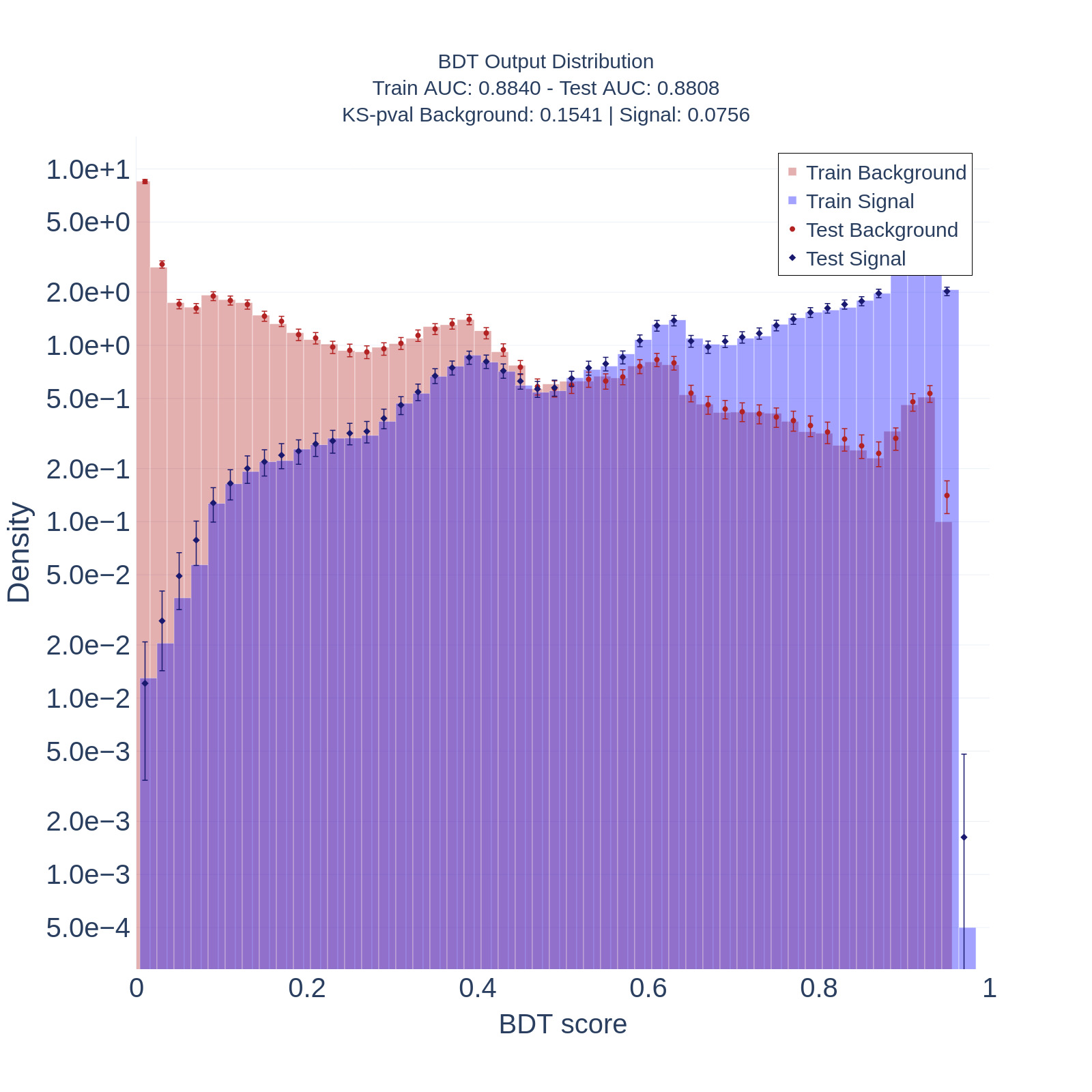}}
\subfigure[]{\includegraphics[width=0.495\textwidth]{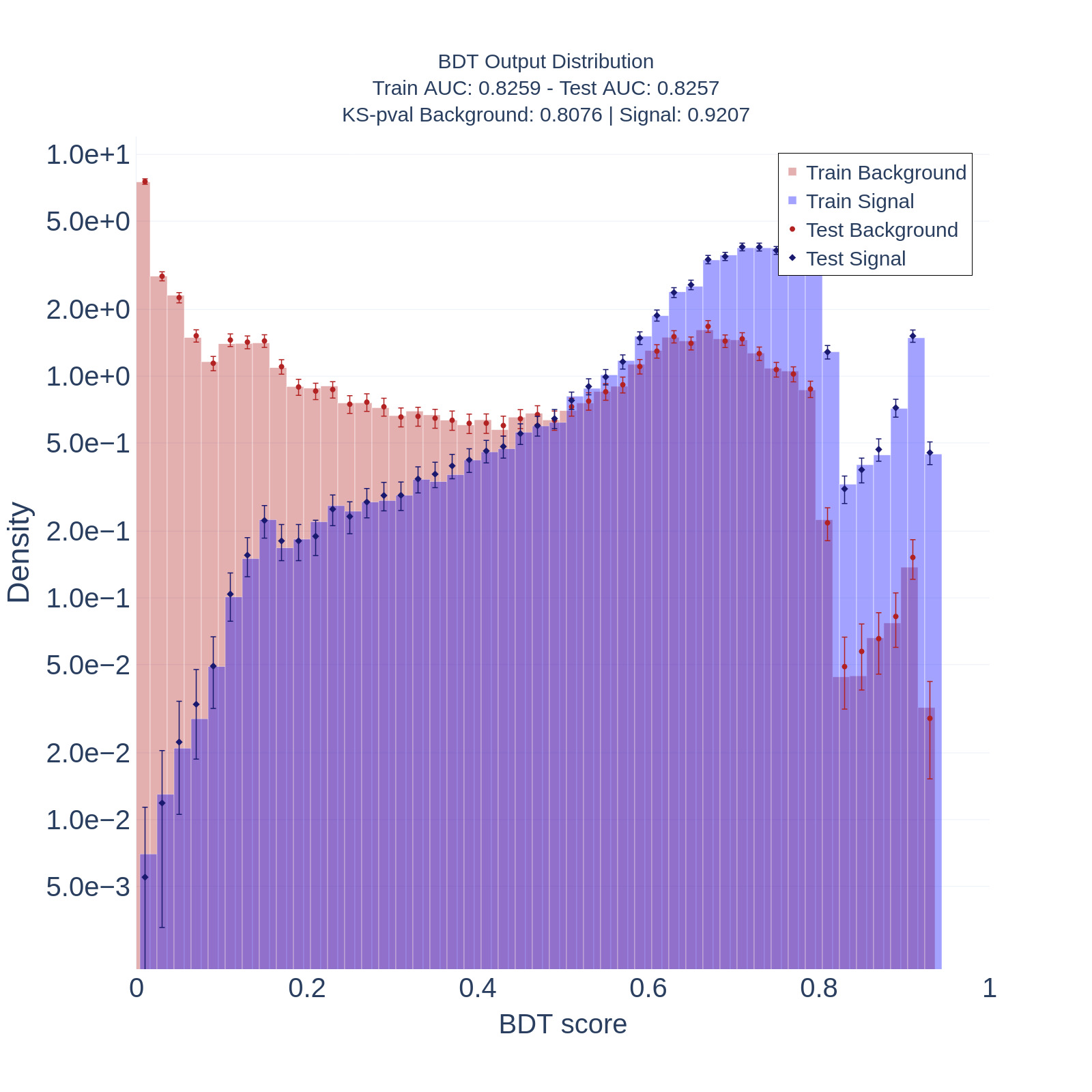}}
\subfigure[]{\includegraphics[width=0.495\textwidth]{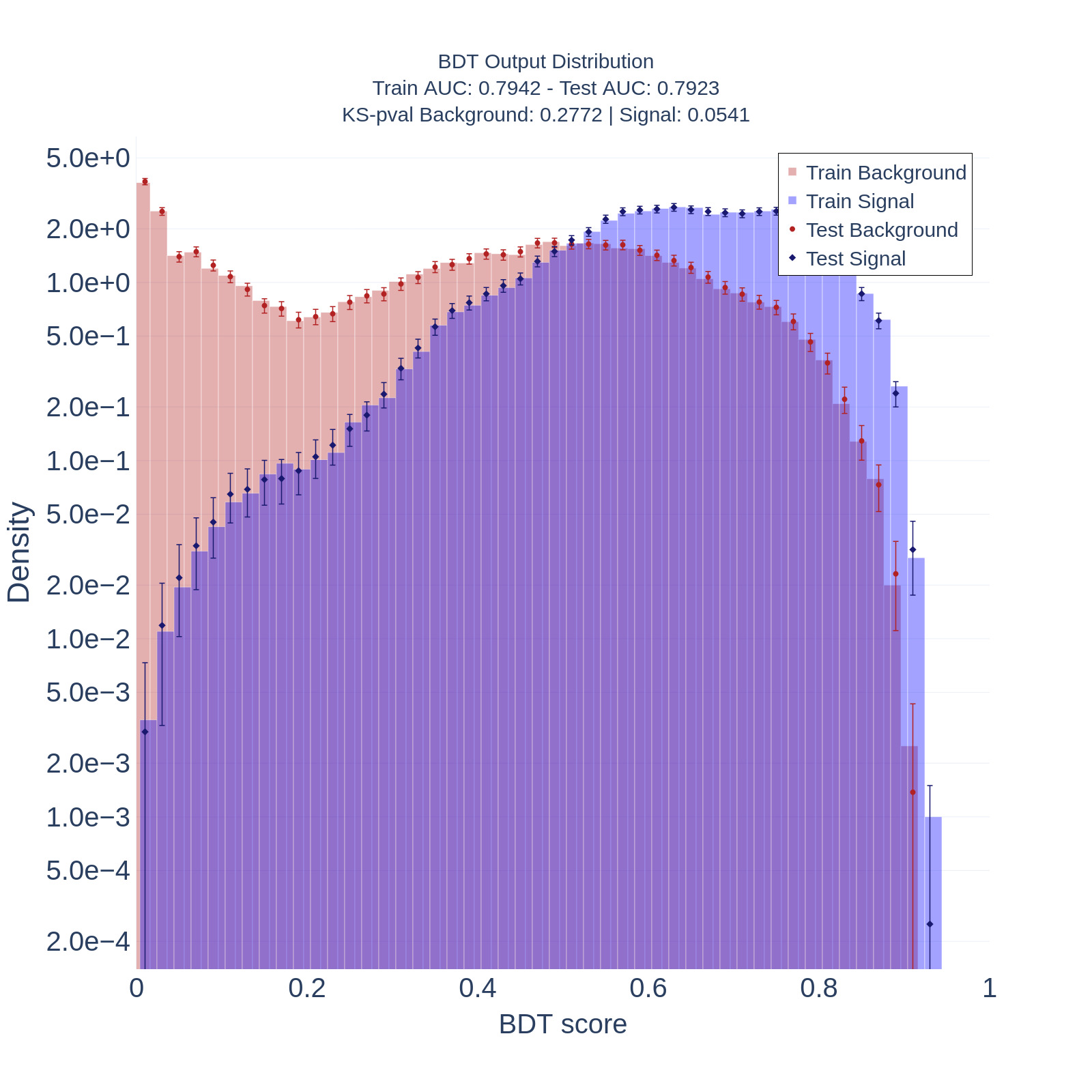}}
\caption{
Distribution of BDT scores for signal and background events across training and testing samples for CEPC: (a) $\sqrt{s}=240$ GeV, (b) $\sqrt{s}=365$ GeV and for FCC-ee: (c) $\sqrt{s}=240$ GeV, (d) $\sqrt{s}=365$ GeV.}
\label{fig:bdt_output}
\end{figure}

A critical irreducible background originates from $Zh$ production, with the Higgs boson decaying to $b\bar{b},\,c\bar{c},\,s\bar{s}$ pairs. These processes share identical final-state kinematics with our signal, posing a unique challenge. As skillfully noted in \cite{Liang:2023wpt}, multivariate techniques present challenges to jet-origin identification. However, we identify viable regions of the parameter space that enhance the cross sections (Fig. \ref{XS_signal}), and therefore we have capable to achieve signal significances at level of $\gtrsim5\sigma$.  Figures~\ref{fig:roc}(a)-(d) displays the Purity vs Efficiency, confirming a moderate but stable discriminatory power against challenging background $e^-e^+\to Zh$. The final classifier achieves area-under-curve (AUC) values of $(a)$ CEPC at $\sqrt{s}=240$ GeV: 0.8055 (training) and 0.8039 (testing), $(b)$ CEPC at $\sqrt{s}=365$ GeV: 0.8840 (training) and 0.8808 (testing), $(c)$ FCC-ee at $\sqrt{s}=240$ GeV: 0.8259 (training) and 0.8257 (testing), and $(d)$ FCC-ee at $\sqrt{s}=365$ GeV: 0.7942 (training) and 0.7923 (testing). 
Meanwhile, we present in Figs.~\ref{fig:bdt_output} (a)-(b) the corresponding BDT output distribution, which shows good agreement between training and testing samples, with Kolmogorov-Smirnov p-values (KS-pval) indicating no overtraining: (a) For CEPC at $\sqrt{s}=240$ GeV we obtain 0.81 for background and 0.14 for signal, (b) CEPC at $\sqrt{s}=365$ GeV 0.15 for background and 0.08 for signal, (c) FCC-ee at $\sqrt{s}=240$ GeV, 0.81 for background and 0.92 for signal, and (d) FCC-ee at $\sqrt{s}=365$ GeV, 0.28 for background and 0.05 for signal. The obtained KS-pval statistic falls within the acceptable range, confirming that the training procedure did not lead to overfitting, despite the inherent difficulty posed by the irreducible $Zh$ background.

Signal and background samples were scaled to the expected event yields, calculated from the integrated luminosity and corresponding cross sections. The BDT selection was optimized separately for each channel to maximize a figure of merit: the signal significance, which is  defined as $\mathcal{N}_S/\sqrt{\mathcal{N}_S+\mathcal{N}_B+(\kappa\cdot \mathcal{N}_B)^2}$, where $\mathcal{N}_S$ and $\mathcal{N}_B$ are the expected numbers of signal and background events, respectively, and $\kappa$ represents the relative systematic uncertainty. The main optimistic experimental systematic uncertainties considered include integrated luminosity ($1\%$), lepton identification efficiencies ($1.5\%$), trigger efficiencies ($1\%$), and momentum scale and resolution ($0.7\%$). Theoretical uncertainties account for QCD scale variations ($2\%$), parton distribution functions ($1.5\%$), and background modeling, particularly for the $Zh$ process ($2\%$). An additional $2\%$ uncertainty is attributed to the multivariate classifier itself. These contributions are combined in quadrature, resulting in a total systematic uncertainty of $4.36\%$. The BDT hyperparameters were optimized using the Optuna framework \cite{akiba2019optuna}\footnote{To facilitate the reproduction of the results reported herein, the hyperparameters employed in the BDT analysis as well as the datasets are provided \href{https://u.pcloud.link/publink/show?code=kZldzE5ZsSsHJPo6iDBPl1pNM01BiVEC4ifV}{\textbf{here}} or upon request.}. 

	\subsection{Signal significance}
	We now turn to present the most important results we found: the signal significance, for which we present two cases, at center-of-mass energies of $\sqrt{s}=240$ and $\sqrt{s}=365$ GeV.
  
	\subsubsection*{CEPC at center-of-mass energy of 240 GeV}
\begin{figure}[!htb]
\begin{center}
	\subfigure[]{\includegraphics[scale=0.21]{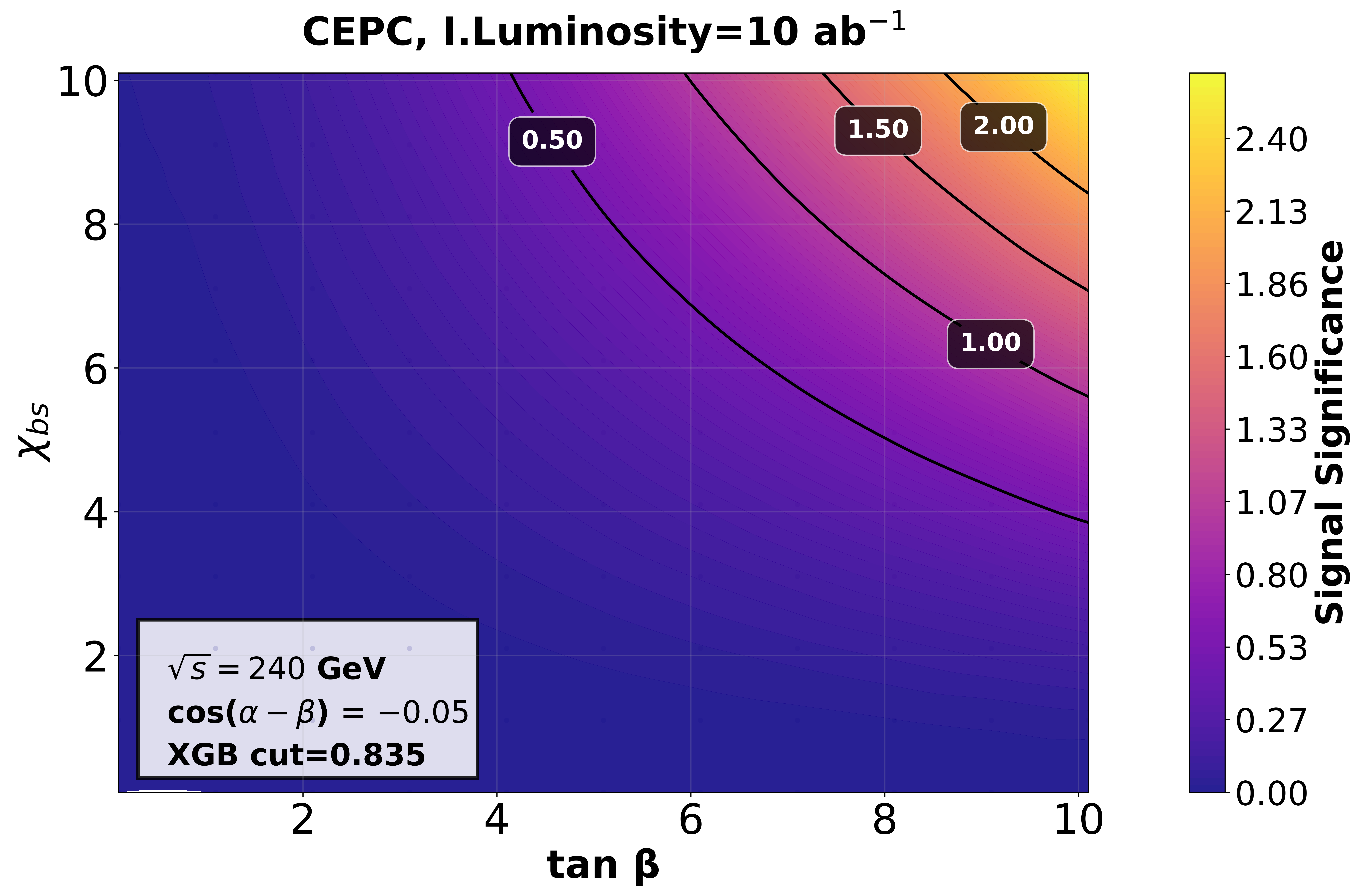}}
    \subfigure[]{\includegraphics[scale=0.13]{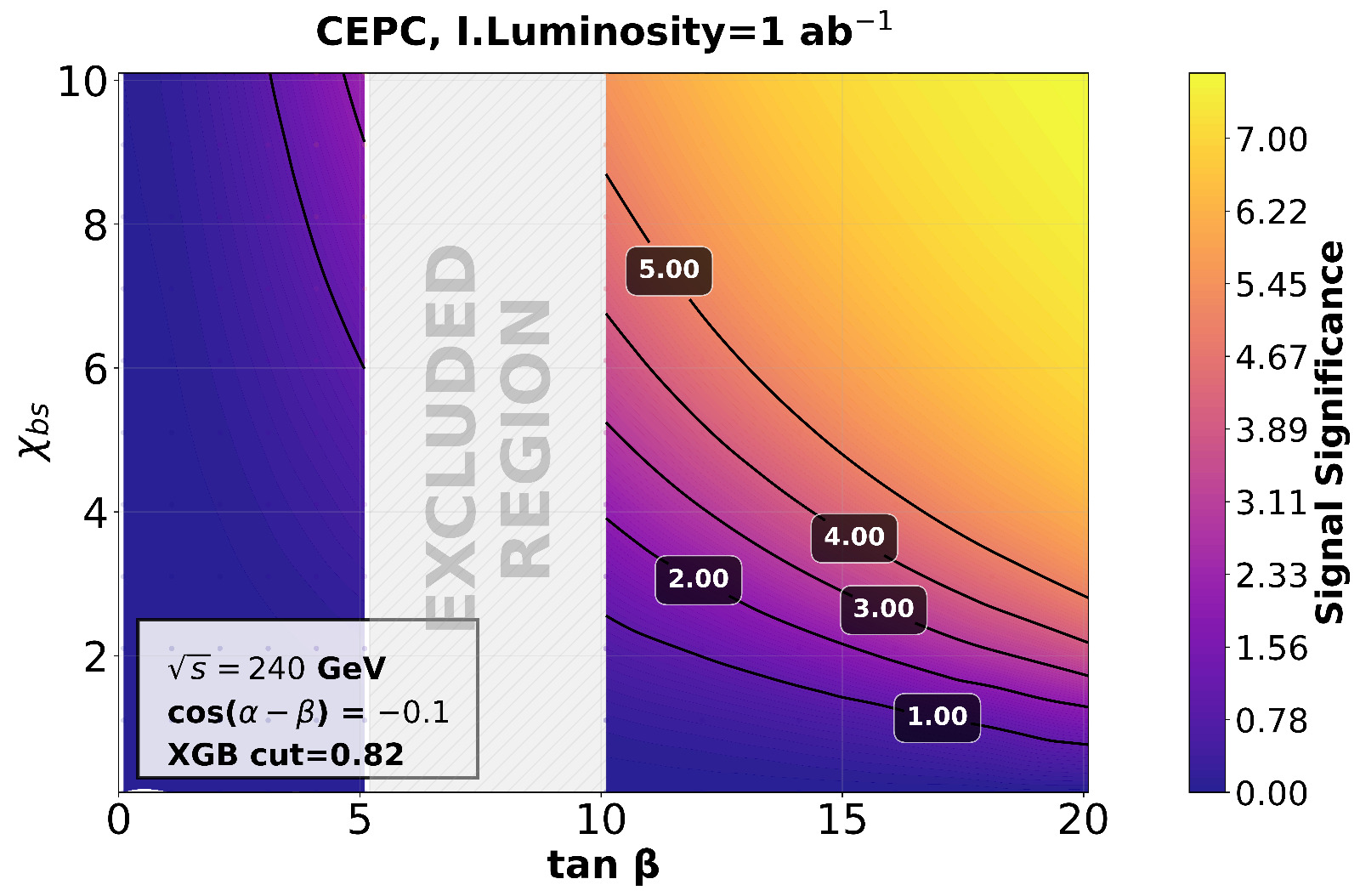}}
 \end{center}
	\caption{Dependence of the signal significance at the CEPC ($\sqrt{s}=240$ GeV) on the flavor-changing parameter $\chi_{bs}$ and $\tan\beta$ under two scenarios: (a) integrated luminosity of $10$ ab$^{-1}$ and $\cos(\alpha-\beta) = -0.05$; (b) integrated luminosity of $1$ ab$^{-1}$ and $\cos(\alpha-\beta) = -0.1$. The significance calculation incorporates a $4\%$ systematic uncertainty.}
 \label{SignificanceCEPC}
\end{figure}

	In Fig.~\ref{SignificanceCEPC} we present the significance of the signal as a function of the flavor-changing parameter $\chi_{bs}$ and $\tan\beta$ for integrated luminosities of 10 ab$^{-1}$ and 1 ab$^{-1}$, which correspond to $\cos(\alpha-\beta)=-0.05$ and $\cos(\alpha-\beta)=-0.1$, respectively. A BDT cut of 0.835 and 0.82 was used to separate the signal from the background.
As shown in Fig. \ref{SignificanceCEPC}(a), this scenario yields exclusion limits at the $2\sigma$ confidence level in a region characterized by high parameter values, namely $\chi_{bs} \sim 10$ and $\tan\beta \sim 10$. This behavior arises from the intrinsically lower cross section of the benchmark with $\cos(\alpha-\beta) = -0.05$, which is approximately one order of magnitude smaller than that corresponding to $\cos(\alpha-\beta) = -0.1$. Such a substantial reduction in production rate directly translates into a diminished signal significance. In contrast, Fig. \ref{SignificanceCEPC}(b) reveals considerably more optimistic prospects. A discovery sensitivity exceeding $5\sigma$ is projected across a broad region of the parameter space, spanning $3 \lesssim \chi_{bs} \lesssim 9$ and $10 \lesssim \tan\beta \lesssim 20$. Remarkably, this sensitivity is already achievable with an integrated luminosity of only 1 ab$^{-1}$, suggesting that the $h \to bs$ decay could be within reach during the early operational phase of CEPC, offering an great opportunity for probing flavor-violating Higgs couplings.

	\subsubsection*{CEPC at a center-of-mass energy of 365 GeV}
\begin{figure}[!htb]
	\begin{center}
		\subfigure[]{\includegraphics[scale=0.13]{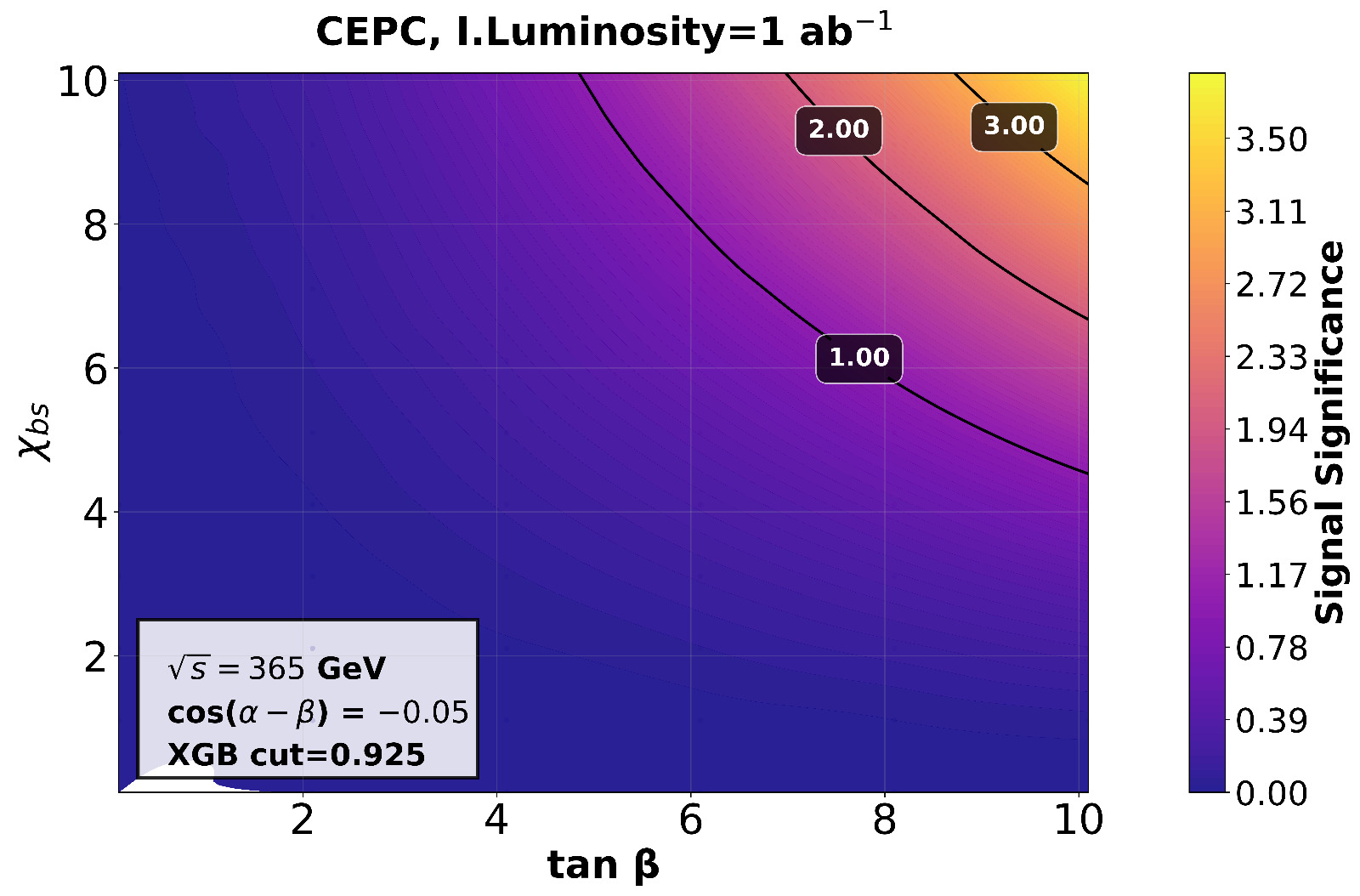}}
        \subfigure[]{\includegraphics[scale=0.13]{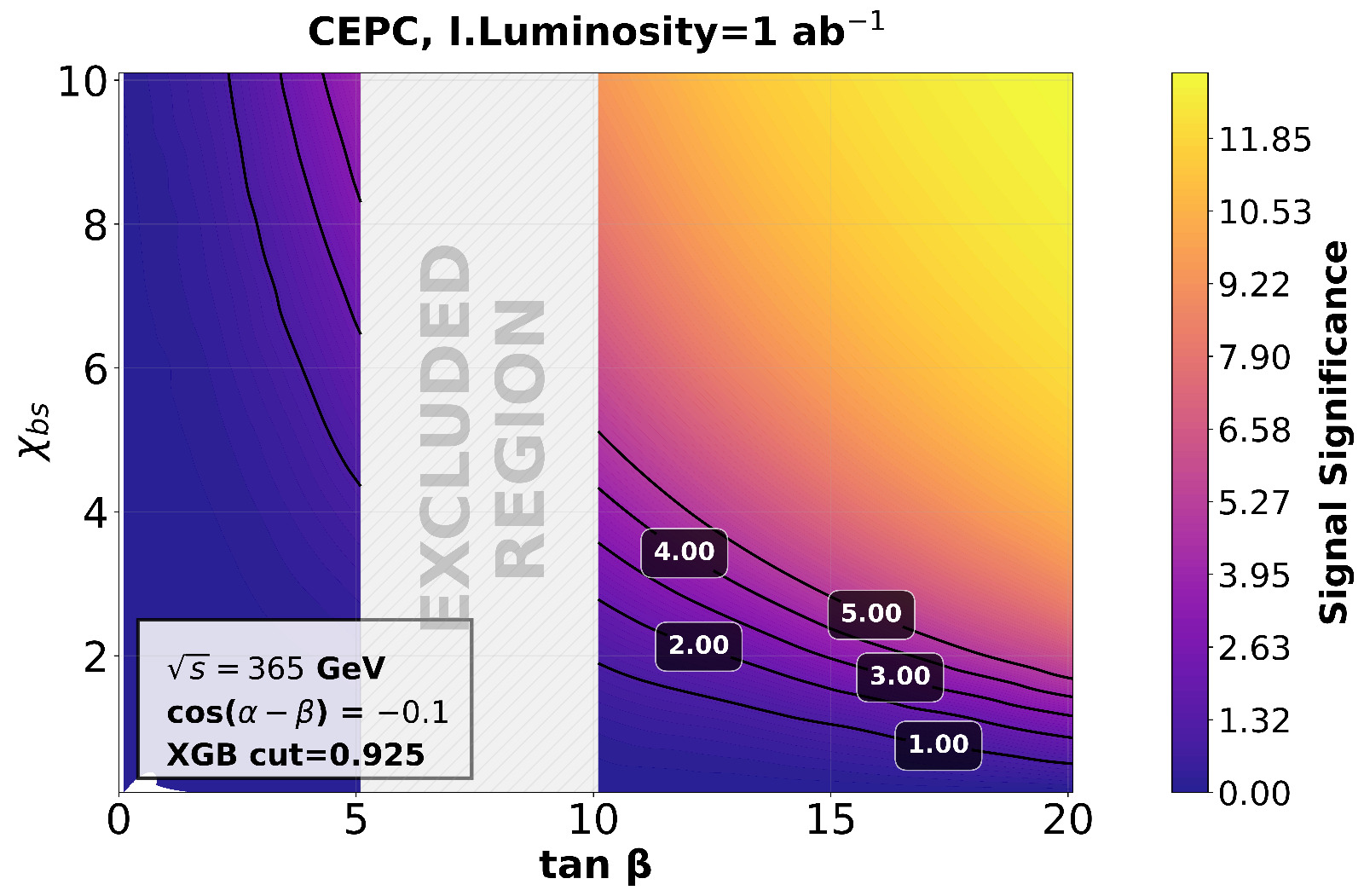}}
	\end{center}
	\caption{Dependence of the signal significance at the CEPC ($\sqrt{s}=365$ GeV) on the flavor-changing parameter $\chi_{bs}$ and $\tan\beta$ under two scenarios: (a) $\cos(\alpha-\beta) = -0.05$; (b) $\cos(\alpha-\beta) = -0.1$. The significance calculation incorporates a $4\%$ systematic uncertainty. }
	\label{Significance365GeV}
\end{figure}

At a center-of-mass energy of 365 GeV, the signal cross section is reduced by a factor of approximately between 0.6-0.7 relative to the 240 GeV case. Additionally, a new background process —$e^-e^+\to t\bar{t}$— becomes kinematically accessible, although its impact on signal significance remains modest. As with the signal, the dominant background processes also experience a reduction in their cross sections, with suppression factors ranging from 0.49 to 0.87 (see Table~\ref{tb:XS-BGD350}). Despite the decrease in the signal cross section, the overall significance improves at $\sqrt{s}=365$ GeV, primarily due to the substantially lower background yields expected at this higher energy. Figure~\ref{Significance365GeV} displays the signal significance as a function of the flavor-violating parameters $\chi_{bs}$, $\tan\beta$ for (a) $\cos(\alpha-\beta)=-0.05$ and (b) $\cos(\alpha-\beta)=-0.1$. In both cases an integrated luminosity of 1 ab$^{-1}$, and BDT cut of 0.925 were considered.
	From Fig.~\ref{Significance365GeV}(b) we observe a potential discovery of the signal process if one considers integrated luminosities $\gtrsim 1$ ab$^{-1}$, $2 \lesssim \chi_{bs} \lesssim 6$ and $10 \lesssim \tan\beta \lesssim 20$. These results, combined with those reported for the $240$ GeV case, present an excellent opportunity for experimental scrutiny. Such validation could lead to an irrefutable discovery of physics beyond the SM. 

\subsubsection*{FCC-ee at center-of-mass energy of 240 GeV}
\begin{figure}[!htb]
\begin{center}
	\subfigure[]{\includegraphics[scale=0.13]{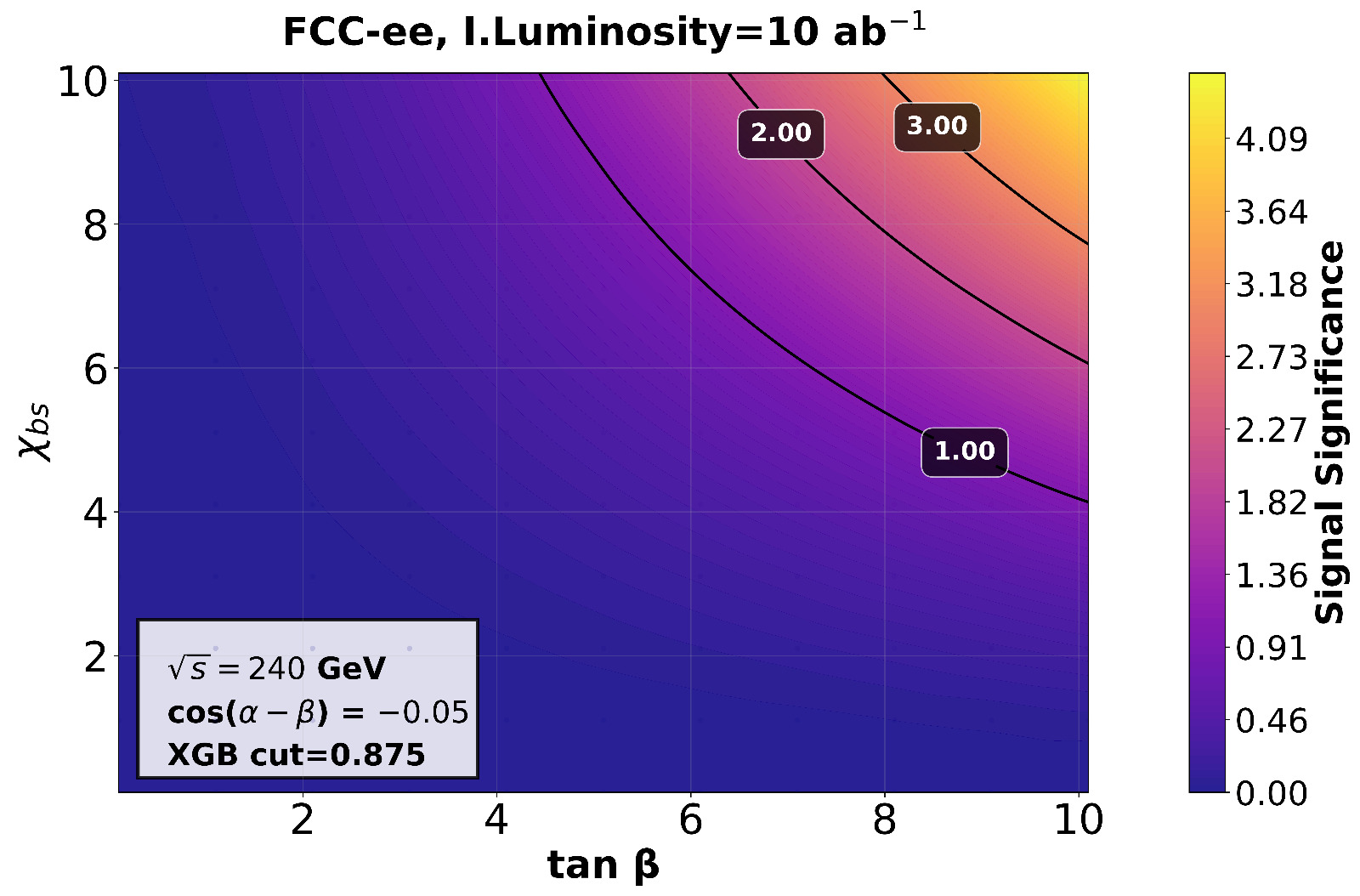}}
    \subfigure[]{\includegraphics[scale=0.13]{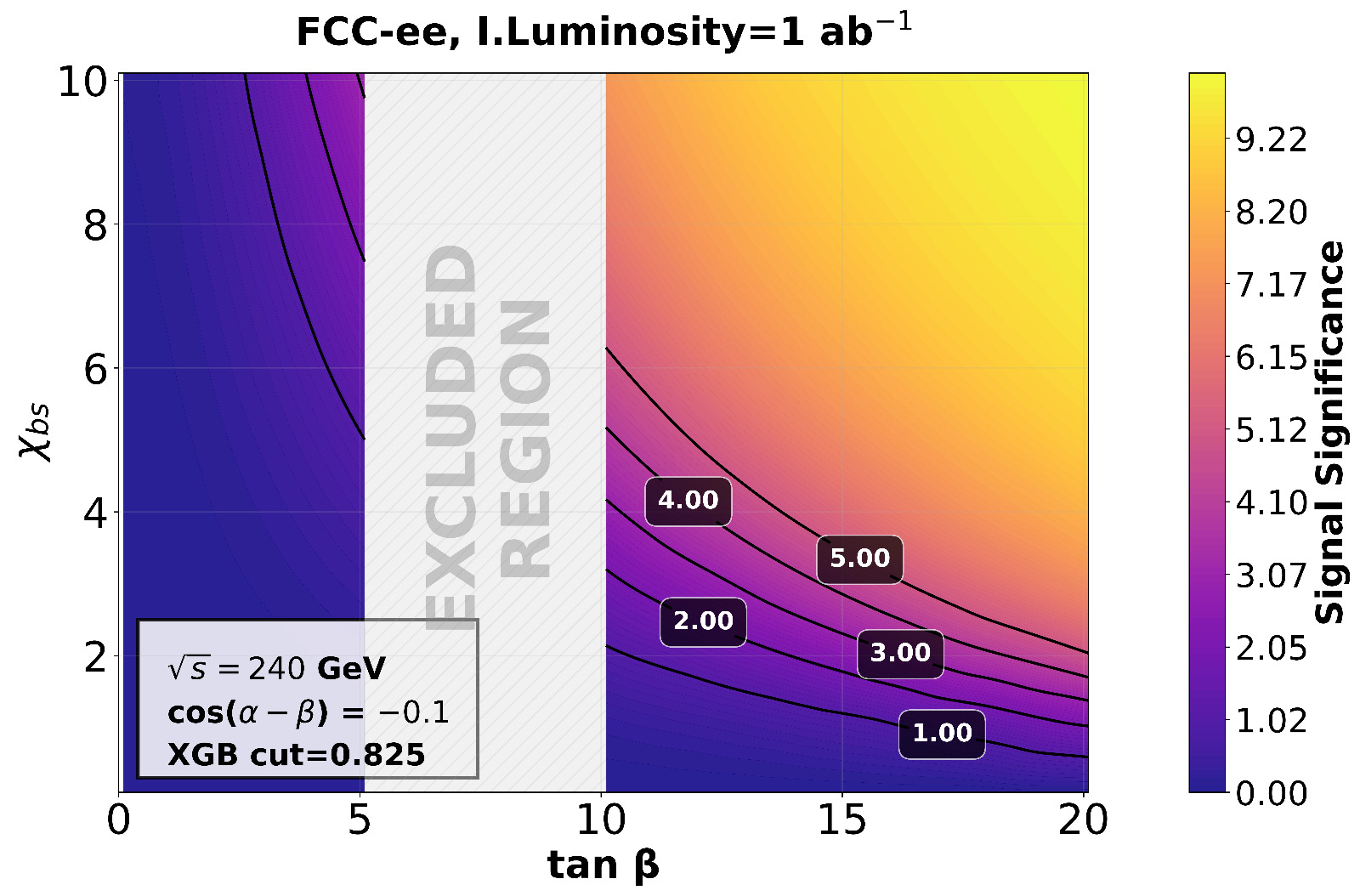}}
 \end{center}
	\caption{Dependence of the signal significance at the FCC-ee ($\sqrt{s}=240$ GeV) on the flavor-changing parameter $\chi_{bs}$ and $\tan\beta$ under two scenarios: (a) integrated luminosity of $10$ ab$^{-1}$ and $\cos(\alpha-\beta) = -0.05$; (b) integrated luminosity of $1$ ab$^{-1}$ and $\cos(\alpha-\beta) = -0.1$. The significance calculation incorporates a $4\%$ systematic uncertainty.}
 \label{SignificanceFCC}
\end{figure}

We now proceed to present the corresponding results for the FCC-ee. To this end, Fig.~\ref{SignificanceFCC} illustrates the signal significance as a function of the flavor-changing parameter $\chi_{bs}$ and $\tan\beta$, for integrated luminosities of $10$ ab$^{-1}$ and $1$ ab$^{-1}$, corresponding to $\cos(\alpha-\beta) = -0.05$ and $\cos(\alpha-\beta) = -0.1$, respectively. A BDT cut of $0.875$ and $0.845$ is applied to isolates the signal from the background.
Analogously to the CEPC case — Fig.~\ref{SignificanceCEPC}(a) — we also derive exclusion limits at the $2\sigma$ confidence level for $\cos(\alpha-\beta) = -0.05$, with $\chi_{bs} \sim 10$ and $\tan\beta \sim 6$, and vice versa (assuming a luminosity of $10$ ab$^{-1}$). As in the CEPC analysis, this behavior stems from the suppression of the cross section by $\cos(\alpha-\beta) = -0.05$ (see Fig.~\ref{XS_signal}(a)). Conversely, $\cos(\alpha-\beta) = -0.1$ accommodates larger $\tan\beta$ values, leading to cross sections roughly one order of magnitude higher. This enhancement in the production rate directly boosts the signal significance by a factor of about $2.5$.
As can be observed in Fig.~\ref{SignificanceFCC}(b), the $\cos(\alpha-\beta) = -0.1$ benchmark presents a considerably more favorable scenario. A signal significance at the level of $\geq 5\sigma$ is projected over a broad region of the parameter space, specifically for $2 \lesssim \chi_{bs} \lesssim 6$ and $10 \lesssim \tan\beta \lesssim 20$. This level of sensitivity is attainable with an integrated luminosity of approximately $1$ ab$^{-1}$, suggesting that the $h \to bs$ decay may be within reach already during the early operational phase of the FCC-ee.

\subsubsection*{FCC-ee at center-of-mass energy of 365 GeV}
\begin{figure}[!htb]
\begin{center}
	\subfigure[]{\includegraphics[scale=0.13]{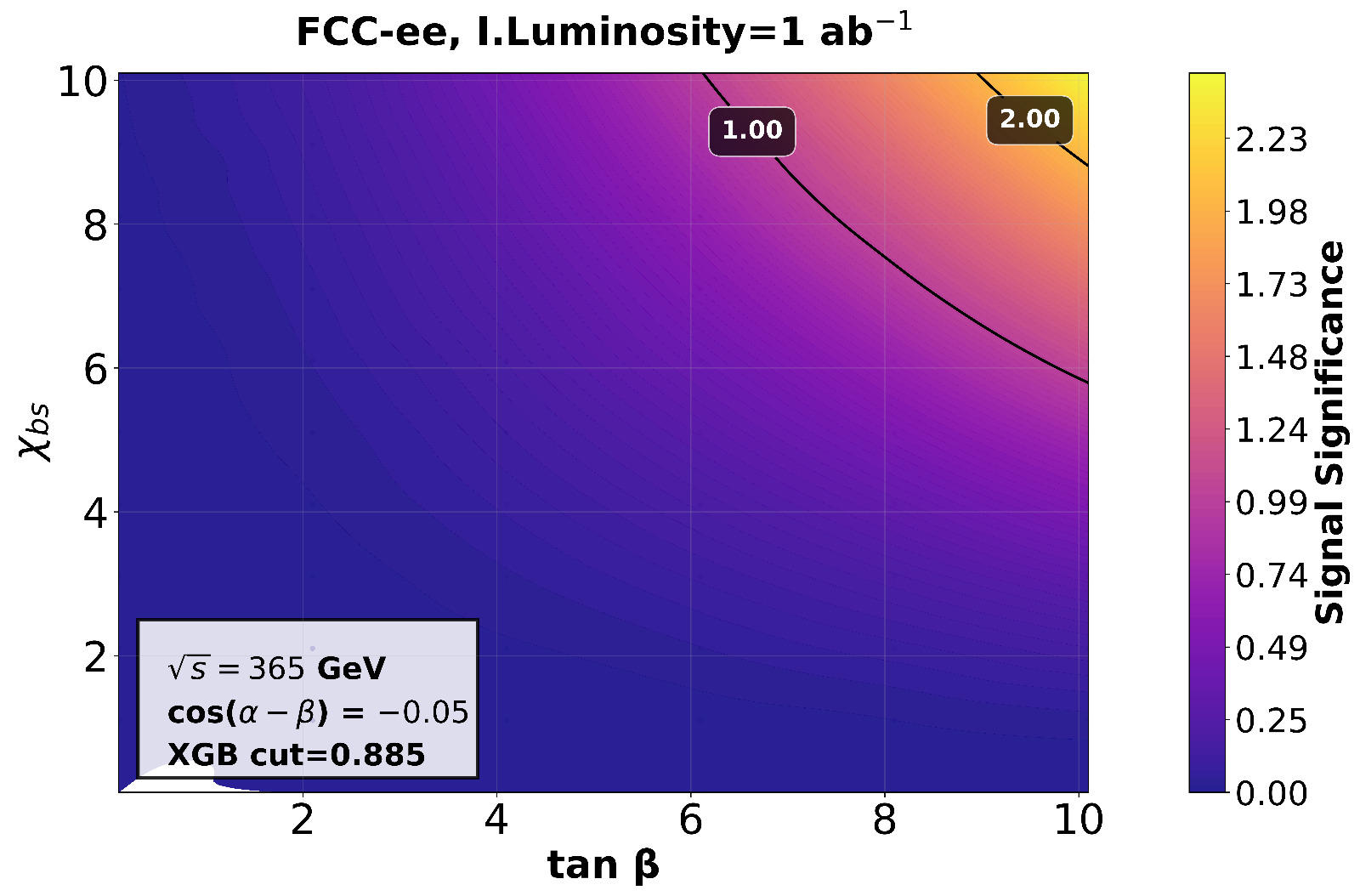}}
    \subfigure[]{\includegraphics[scale=0.13]{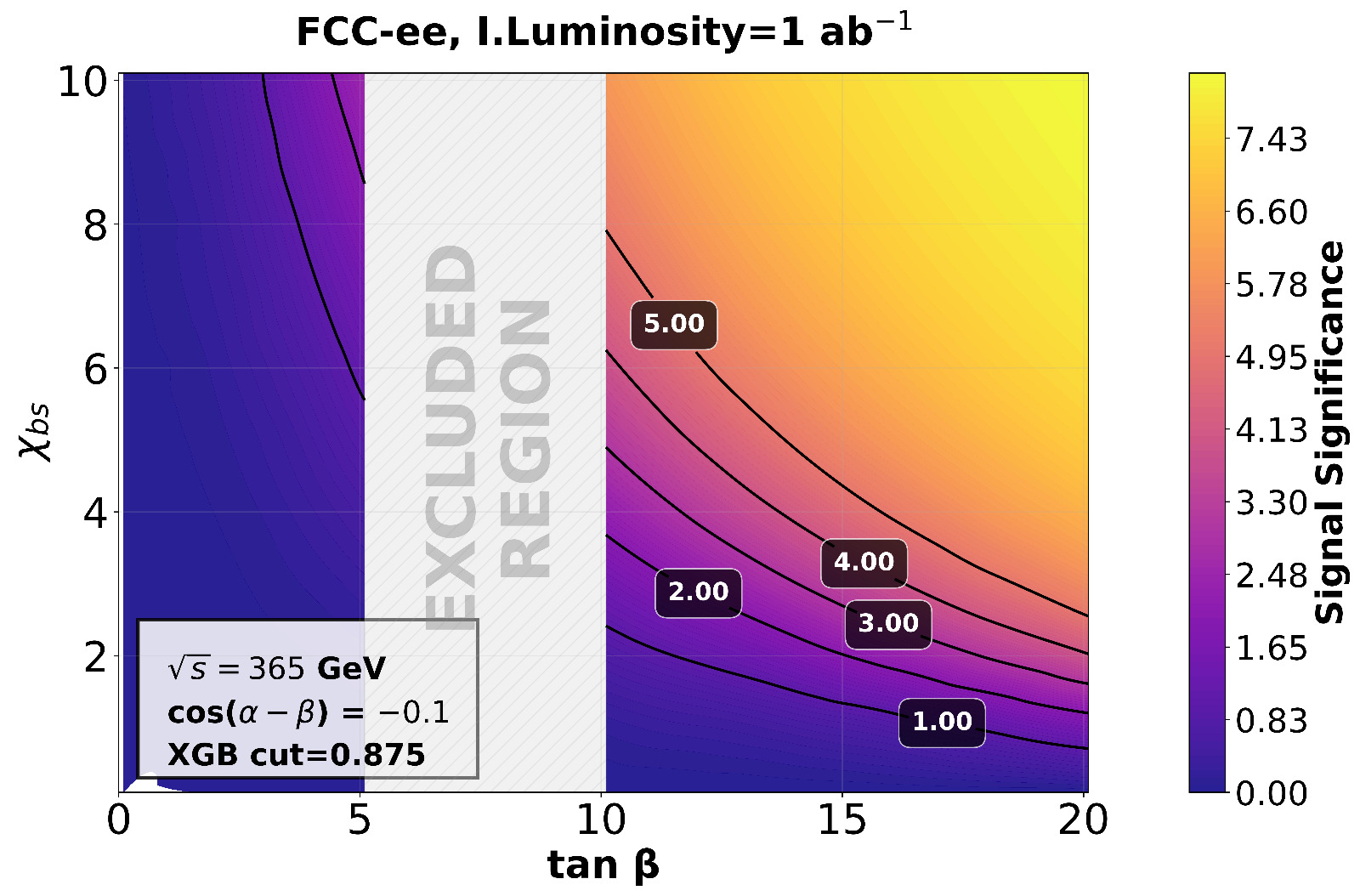}}
 \end{center}
	\caption{Dependence of the signal significance at the FCC-ee ($\sqrt{s}=365$ GeV) on the flavor-changing parameter $\chi_{bs}$ and $\tan\beta$ under two scenarios: (a) integrated luminosity of $1$ ab$^{-1}$ and $\cos(\alpha-\beta) = -0.05$; (b) integrated luminosity of $1$ ab$^{-1}$ and $\cos(\alpha-\beta) = -0.1$. The significance calculation incorporates a $4\%$ systematic uncertainty.}
 \label{SignificanceFCC365}
\end{figure}

Finally, we present in Fig.~\ref{SignificanceFCC365} the results obtained for FCC-ee at $\sqrt{s}=365$ GeV.
The results found indicate the FCC-ee at $\sqrt{s}=365$ GeV and an integrated luminosity of 1 ab$^{-1}$ excludes a significant region on the plane ($\tan\beta-\chi_{bs}$) if one considers the case $\cos(\alpha-\beta)=-0.05$. Such a value is close to the decoupling limit, entering to the SM regimen. In contrast, when $\cos(\alpha-\beta)=-0.1$ is considered, the significance takes values at level of $\geq5\sigma$ for the ranges $3\lesssim \chi_{bs}\lesssim 8$ and $10 \lesssim \tan\beta \lesssim 20$, and an integrated luminosity of 1 ab$^{-1}$.      
	\section{Conclusions}\label{SecV}

In this work, we have systematically investigated the discovery potential of the flavor-changing Higgs decay \(h \to bs\) within the framework of the Two-Higgs-Doublet Model Type~III (2HDM-III), focusing on the Higgs-strahlung process \(e^+e^- \to Zh\) at future circular lepton colliders — namely, the CEPC and the FCC-ee — at center-of-mass energies of \(\sqrt{s} = 240\)~GeV and \(365\)~GeV.

A comprehensive exploration of the model parameter space was performed, incorporating the most stringent constraints from LHC Higgs signal strengths, lepton-flavor-violating decays, and the rare decay \(B_s \to \mu^+\mu^-\). The latter plays a critical role in bounding the flavor-changing parameter \(\chi_{bs}\). We identified viable regions where the branching ratio \(\mathcal{BR}(h \to bs)\) reaches \(\mathcal{O}(10^{-3})\), i.e., several orders of magnitude above the Standard Model prediction, while remaining consistent with all current experimental bounds.

Using a full simulation chain — from parton-level event generation with \textsc{MadGraph5}, showering and hadronization with \textsc{Pythia8}, to detector response emulation with \textsc{Delphes~3} — we implemented a multivariate analysis based on Boosted Decision Trees (BDT) to isolate the signal from irreducible backgrounds, particularly \(Zh\) production with \(h \to b\bar{b}, c\bar{c}, s\bar{s}\). The BDT classifiers achieved solid and stable performance for both CEPC and FCC-ee, with AUC values in the range \(\sim 0.79\)–\(0.88\) and Kolmogorov-Smirnov \(p\)-values confirming the absence of overtraining. This validates the robustness of our projected significances for both colliders.

The main results can be summarized as follows:
\begin{itemize}
    \item \textbf{At CEPC and FCC-ee with \(\sqrt{s}=240\)~GeV}, the benchmark scenario with \(\cos(\alpha-\beta) = -0.1\) and an integrated luminosity of \(1~\text{ab}^{-1}\) yields a \(5\sigma\) discovery reach over broad regions of the parameter space: for CEPC, \(3 \lesssim \chi_{bs} \lesssim 9\) and \(10 \lesssim \tan\beta \lesssim 20\); for FCC-ee, \(2 \lesssim \chi_{bs} \lesssim 6\) and \(10 \lesssim \tan\beta \lesssim 20\). These results indicate that the \(h \to bs\) decay could be accessible already during the early operational phase of both colliders.
    
    \item \textbf{At \(\sqrt{s}=365\)~GeV}, although the signal cross section is reduced, the even stronger suppression of dominant backgrounds — including \(t\bar{t}\) production — allows a \(5\sigma\) sensitivity to also be achieved with \(1~\text{ab}^{-1}\) for both CEPC and FCC-ee, particularly in the \(\cos(\alpha-\beta) = -0.1\) scenario. This provides a valuable cross-check for any potential signal observed at \(240\)~GeV.
    
    \item For the more conservative scenario \(\cos(\alpha-\beta) = -0.05\), which lies closer to the decoupling limit, exclusion limits at \(2\sigma\) confidence level are projected for both colliders, with \(\chi_{bs} \sim 10\) and \(\tan\beta \sim 6\)–\(10\). By considering this scenario (and an integrated luminosity of 10 ab$^{-1}$), we found a lower limit for $\mathcal{BR}(h\to bs)\leq\mathcal{O}(10^{-3})$, which is potentially sensitive to be measured at CEPC and FCC-ee.
\end{itemize}

In conclusion, this work shows that future circular electron-positron colliders — both CEPC and FCC-ee — offer a unique and powerful avenue to probe flavor-changing Higgs couplings. The decay \(h \to bs\), although challenging, emerges as a promising channel to uncover signatures of physics beyond the Standard Model. We encourage the experimental community to further explore this channel, as it could open a new window into the flavor structure of the Higgs sector in the coming decades.
	\section*{Acknowledgments}
The work of Marco A. Arroyo-Ure\~na and T. Valencia-P\'erez is supported by ``Estancias Posdoctorales por M\'exico (SECIHTI)'' and ``Sistema Nacional de Investigadoras e Investigadores'' (SNII-SECIHTI). We thank Haydee Hernández-Arellano for her valuable comments that improved this work and Enrique Díaz for his suggestions on the writing and clarity of the manuscript. T.V.P. acknowledges support from the UNAM project PAPIIT IN111224 and the SECIHTI project CBF2023-2024-548.	
	\appendix
	\section{Explicit form of the $\mathcal{O}_{f}$ matrix}\label{Omatrix}
The $\mathcal{O}_f$ matrix that diagonalize the fermion mass matrix via $V_f=\mathcal{O}_fP_f$, with $P_f=\text{diag}\{e^{i\alpha_f},\,e^{i\beta_f},\,1\}$, is given by	

		\begin{equation}
			\mathcal{O}_{f}=\left(\begin{array}{ccc}
				\sqrt{\dfrac{m_{f_2}m_{f_3}(A-m_{f_1})}{A(m_{f_2}-m_{f_1})(m_{f_3}-m_{f_1})}} & \sqrt{\dfrac{m_{f_1}m_{f_3}(m_{f_2}-A)}{A(m_{f_2}-m_{f_1})(m_{f_3}-m_{f_2})}} & \sqrt{\dfrac{m_{f_1}m_{f_3}(A-m_{f_3})}{A(m_{f_3}-m_{f_1})(m_{f_3}-m_{f_2})}}\\
				-\sqrt{\dfrac{m_{f_1}(m_{f_1}-A)}{(m_{f_2}-m_{f_1})(m_{f_3}-m_{f_1})}} & \sqrt{\dfrac{m_{f_2}(A-m_{f_2})}{(m_{f_2}-m_{f_1})(m_{f_3}-m_{f_2})}} & \sqrt{\dfrac{m_{f_3}(m_{f_2}-A)}{(m_{f_2}-m_{f_1})(m_{f_3}-m_{f_2})}}\\
				\sqrt{\dfrac{m_{f_1}(A-m_{f_2})(A-m_{f_3})}{A(m_{f_2}-m_{f_1})(m_{f_3}-m_{f_1})}} & -\sqrt{\dfrac{m_{f_2}(A-m_{f_1})(m_{f_3}-A)}{A(m_{f_2}-m_{f_1})(m_{f_3}-m_{f_2})}} & \sqrt{\dfrac{m_{f_3}(A-m_{f_1})(A-m_{f_2})}{A(m_{f_3}-m_{f_1})(m_{f_3}-m_{f_2})}}
			\end{array}\right),
		\end{equation}

where $m_{f_i}$ ($i=1,\,2,\,3$) are the physical fermion masses of the three generations.	

\bibliographystyle{JHEP}
\bibliography{aipsamp}

@article{MEG:2016leq,
    author = "Baldini, A. M. and others",
    collaboration = "MEG",
    title = "{Search for the lepton flavour violating decay $\mu ^+ \rightarrow \mathrm {e}^+ \gamma $ with the full dataset of the MEG experiment}",
    eprint = "1605.05081",
    archivePrefix = "arXiv",
    primaryClass = "hep-ex",
    doi = "10.1140/epjc/s10052-016-4271-x",
    journal = "Eur. Phys. J. C",
    volume = "76",
    number = "8",
    pages = "434",
    year = "2016"
}

@article{BaBar:2009hkt,
    author = "Aubert, Bernard and others",
    collaboration = "BaBar",
    title = "{Searches for Lepton Flavor Violation in the Decays $\tau^{\pm}\to e^{\pm} \gamma$ and $\tau^{\pm}\to\mu^{\pm} \gamma$}",
    eprint = "0908.2381",
    archivePrefix = "arXiv",
    primaryClass = "hep-ex",
    reportNumber = "SLAC-PUB-13753, BABAR-PUB-09-026",
    doi = "10.1103/PhysRevLett.104.021802",
    journal = "Phys. Rev. Lett.",
    volume = "104",
    pages = "021802",
    year = "2010"
}

@article{Belle:2021ysv, 
    author = "Abdesselam, A. and others",
    collaboration = "Belle",
    title = "{Search for lepton-flavor-violating tau-lepton decays to $\ell\gamma$ at Belle}",
    eprint = "2103.12994",
    archivePrefix = "arXiv",
    primaryClass = "hep-ex",
    doi = "10.1007/JHEP10(2021)019",
    journal = "JHEP",
    volume = "10",
    pages = "19",
    year = "2021"
}

@book{Schopper:2009zz,
    author = "Schopper, Herwig",
    title = "{LEP - The Lord of the Collider Rings at CERN 1980-2000. The Making, Operation and Legacy of the World's Largest Scientific Instrument}",
    doi = "10.1007/978-3-540-89301-1",
    isbn = "978-3-540-89300-4, 978-3-662-50145-0, 978-3-540-89301-1",
    publisher = "Springer",
    year = "2009"
}

@inproceedings{akiba2019optuna,
  title={Optuna: A Next-generation Hyperparameter Optimization Framework},
  author={Akiba, Takuya and Sano, Shotaro and Yanase, Toshihiko and Ohta, Takeru and Koyama, Masanori},
  booktitle={Proceedings of the 25th ACM SIGKDD International Conference on Knowledge Discovery \& Data Mining},
  pages={2623--2631},
  year={2019}
}

@article{Liang:2023wpt,
    author = "Liang, Hao and Zhu, Yongfeng and Wang, Yuexin and Che, Yuzhi and Zhou, Chen and Qu, Huilin and Ruan, Manqi",
    title = "{Jet-Origin Identification and Its Application at an Electron-Positron Higgs Factory}",
    eprint = "2310.03440",
    archivePrefix = "arXiv",
    primaryClass = "hep-ex",
    doi = "10.1103/PhysRevLett.132.221802",
    journal = "Phys. Rev. Lett.",
    volume = "132",
    number = "22",
    pages = "221802",
    year = "2024"
}

@inproceedings{Chen:2016:XST:2939672.2939785,  
author = {Chen, Tianqi and Guestrin, Carlos},  
title = "{XGBoost}: A Scalable Tree Boosting System",  
booktitle = {Proc. 22nd ACM SIGKDD Int. Conf. on Knowledge Discovery and Data Mining},  
series = {KDD '16},  year = {2016},  
isbn = {978-1-4503-4232-2},  
location = {San Francisco, California, USA},  
pages = {785},  
numpages = {10},  
doi = {10.1145/2939672.2939785},  
acmid = {2939785},  
publisher = {ACM},  
address = {New York, NY, USA}, 
}

@book{Book:bdt,
author      = "L., Breiman and others",
title       = "{Classification and regression trees}",
publisher   = "Wadsworth international group",
year        = "1984",
address     = "California, USA",
}

@inproceedings{Maiani:2013fpa,
    author = "Maiani, Luciano",
    title = "{The GIM Mechanism: origin, predictions and recent uses}",
    booktitle = "{48th Rencontres de Moriond on Electroweak Interactions and Unified Theories}",
    eprint = "1303.6154",
    archivePrefix = "arXiv",
    primaryClass = "hep-ph",
    pages = "3--16",
    year = "2013"
}

@article{Apollinari:2015wtw,
    author = {Apollinari, G. and Br{\"u}ning, O. and Nakamoto, T. and Rossi, Lucio},
    editor = {Apollinari, G and B{\'e}jar Alonso, I and Br{\"u}ning, O and Lamont, M and Rossi, L},
    title = "{High Luminosity Large Hadron Collider HL-LHC}",
    eprint = "1705.08830",
    archivePrefix = "arXiv",
    primaryClass = "physics.acc-ph",
    reportNumber = "FERMILAB-PUB-15-699-TD",
    doi = "10.5170/CERN-2015-005.1",
    journal = "CERN Yellow Rep.",
    number = "5",
    pages = "1--19",
    year = "2015"
}

@article{Arkani-Hamed:2015vfh,
    author = "Arkani-Hamed, Nima and Han, Tao and Mangano, Michelangelo and Wang, Lian-Tao",
    title = "{Physics opportunities of a 100 TeV proton{\textendash}proton collider}",
    eprint = "1511.06495",
    archivePrefix = "arXiv",
    primaryClass = "hep-ph",
    reportNumber = "PITT-PACC-1515, CERN-PH-TH-2015-259",
    doi = "10.1016/j.physrep.2016.07.004",
    journal = "Phys. Rept.",
    volume = "652",
    pages = "1--49",
    year = "2016"
}

@article{Cepeda:2019klc,
    author = "Cepeda, M. and others",
    editor = "Dainese, Andrea and Mangano, Michelangelo and Meyer, Andreas B. and Nisati, Aleandro and Salam, Gavin and Vesterinen, Mika Anton",
    title = "{Report from Working Group 2}: {Higgs Physics at the HL-LHC and HE-LHC}",
    eprint = "1902.00134",
    archivePrefix = "arXiv",
    primaryClass = "hep-ph",
    reportNumber = "CERN-LPCC-2018-04",
    doi = "10.23731/CYRM-2019-007.221",
    journal = "CERN Yellow Rep. Monogr.",
    volume = "7",
    pages = "221--584",
    year = "2019"
}

@article{Degrande:2011ua,
    author = "Degrande, Celine and Duhr, Claude and Fuks, Benjamin and Grellscheid, David and Mattelaer, Olivier and Reiter, Thomas",
    title = "{UFO - The Universal FeynRules Output}",
    eprint = "1108.2040",
    archivePrefix = "arXiv",
    primaryClass = "hep-ph",
    reportNumber = "CP3-11-25, IPHC-PHENO-11-04, IPPP-11-39, DCPT-11-78, MPP-2011-68",
    doi = "10.1016/j.cpc.2012.01.022",
    journal = "Comput. Phys. Commun.",
    volume = "183",
    pages = "1201--1214",
    year = "2012"
}

@article{CMS:2022mgd,
    author = "Tumasyan, Armen and others",
    collaboration = "CMS",
    title = "{Measurement of the B$^0_\mathrm{S}$$\to$$\mu^+\mu^-$ decay properties and search for the B$^0$$\to$$\mu^+\mu^-$ decay in proton-proton collisions at $\sqrt{s}$ = 13 TeV}",
    eprint = "2212.10311",
    archivePrefix = "arXiv",
    primaryClass = "hep-ex",
    reportNumber = "CMS-BPH-21-006, CERN-EP-2022-270",
    doi = "10.1016/j.physletb.2023.137955",
    journal = "Phys. Lett. B",
    volume = "842",
    pages = "137955",
    year = "2023"
}

@article{Arroyo-Urena:2013cyf,
    author = "Arroyo-Ure{\~n}a, Marco A. and Diaz-Cruz, J. Lorenzo and D{\'\i}az, Enrique and Orduz-Ducuara, Javier A.",
    title = "{Flavor violating Higgs signals in the Texturized Two-Higgs Doublet Model (THDM-Tx)}",
    eprint = "1306.2343",
    archivePrefix = "arXiv",
    primaryClass = "hep-ph",
    doi = "10.1088/1674-1137/40/12/123103",
    journal = "Chin. Phys. C",
    volume = "40",
    number = "12",
    pages = "123103",
    year = "2016"
}

@article{Diaz-Cruz:2004wsi,
    author = "Diaz-Cruz, J. L. and Noriega-Papaqui, R. and Rosado, A.",
    title = "{Mass matrix ansatz and lepton flavor violation in the THDM-III}",
    eprint = "hep-ph/0401194",
    archivePrefix = "arXiv",
    reportNumber = "FCFM-BUAP-HEP-04-01, IFUAP-HEP-04-01",
    doi = "10.1103/PhysRevD.69.095002",
    journal = "Phys. Rev. D",
    volume = "69",
    pages = "095002",
    year = "2004"
}

@article{LorenzoDiaz-Cruz:2019imm,
    author = "Lorenzo D{\'\i}az-Cruz, J.",
    title = "{The Higgs profile in the standard model and beyond}",
    eprint = "1904.06878",
    archivePrefix = "arXiv",
    primaryClass = "hep-ph",
    reportNumber = "CIFFU-19-03",
    doi = "10.31349/RevMexFis.65.419",
    journal = "Rev. Mex. Fis.",
    volume = "65",
    number = "5",
    pages = "419--439",
    year = "2019"
}

@article{Fritzsch:1995nx,
    author = "Fritzsch, Harald and Xing, Zhi-zhong",
    title = "{A Symmetry pattern of maximal CP violation and a determination of the unitarity triangle}",
    eprint = "hep-ph/9502297",
    archivePrefix = "arXiv",
    reportNumber = "MPI-PHT-95-08",
    doi = "10.1016/0370-2693(95)00545-V",
    journal = "Phys. Lett. B",
    volume = "353",
    pages = "114--118",
    year = "1995"
}

@article{Branco:1999nb,
    author = "Branco, G. C. and Emmanuel-Costa, D. and Gonzalez Felipe, R.",
    title = "{Texture zeros and weak basis transformations}",
    eprint = "hep-ph/9911418",
    archivePrefix = "arXiv",
    reportNumber = "FISIST-19-99-CFIF",
    doi = "10.1016/S0370-2693(00)00193-3",
    journal = "Phys. Lett. B",
    volume = "477",
    pages = "147--155",
    year = "2000"
}

@article{Kamenik:2023hvi,
    author = "Kamenik, Jernej F. and Korajac, Arman and Szewc, Manuel and Tammaro, Michele and Zupan, Jure",
    title = "{Flavor-violating Higgs and Z boson decays at a future circular lepton collider}",
    eprint = "2306.17520",
    archivePrefix = "arXiv",
    primaryClass = "hep-ph",
    doi = "10.1103/PhysRevD.109.L011301",
    journal = "Phys. Rev. D",
    volume = "109",
    number = "1",
    pages = "L011301",
    year = "2024"
}

@article{Ai:2024nmn,
    author = "Ai, Xiaocong and others",
    title = "{Flavor Physics at CEPC: a General Perspective}",
    eprint = "2412.19743",
    archivePrefix = "arXiv",
    primaryClass = "hep-ex",
    month = "12",
    year = "2024"
}

@article{Farrera:2020bon,
    author = "Farrera, Carlos M. and Granados-Gonz\'alez, Alejandro and Novales-S\'anchez, H\'ector and Toscano, J. Jes\'us",
    title = "{Quark-flavor-changing Higgs decays from a universal extra dimension}",
    eprint = "2003.05571",
    archivePrefix = "arXiv",
    primaryClass = "hep-ph",
    doi = "10.1142/S0217751X20501419",
    journal = "Int. J. Mod. Phys. A",
    volume = "35",
    number = "24",
    pages = "2050141",
    year = "2020"
}

@article{Coadou:2022nsh,
    author = {Coadou, Yann},
    title = {Boosted decision trees},
    eprint = {2206.09645},
    archivePrefix = {arXiv},
    primaryClass = {physics.data-an},
    doi = {10.1142/9789811234033\_0002},
    month = {3},
    year = {2022}
}

@book{Hastie:2009itz,
    author = "Hastie, Trevor and Tibshirani, Robert and Friedman, Jerome",
    title = "{The Elements of Statistical Learning}",
    doi = "10.1007/978-0-387-84858-7",
    isbn = "978-0-387-84857-0, 978-0-387-84858-7",
    publisher = "Springer",
    year = "2009"
}

@article{ILCInternationalDevelopmentTeam:2022izu,
    author = "Aryshev, Alexander and others",
    collaboration = "ILC International Development Team",
    title = "{The International Linear Collider: Report to Snowmass 2021}",
    eprint = "2203.07622",
    archivePrefix = "arXiv",
    primaryClass = "physics.acc-ph",
    reportNumber = "DESY-22-045, IFT-UAM/CSIC-22-028, KEK Preprint 2021-61, IFT--UAM/CSIC--22-028, KEK Preprint 2021-61,
  PNNL-SA-160884, SLAC-PUB-17662, FERMILAB-FN-1171-PPD-QIS-SCD-TD, PNNL-SA-160884",
    month = "3",
    year = "2022"
}

@article{FCC:2018evy,
    author = "Abada, A. and others",
    collaboration = "FCC",
    title = "{FCC-ee: The Lepton Collider}: {Future Circular Collider Conceptual Design Report Volume 2}",
    reportNumber = "CERN-ACC-2018-0057",
    doi = "10.1140/epjst/e2019-900045-4",
    journal = "Eur. Phys. J. ST",
    volume = "228",
    number = "2",
    pages = "261--623",
    year = "2019"
}

@article{Higgs:1964pj,
    author = "Higgs, Peter W.",
    editor = "Taylor, J. C.",
    title = "{Broken Symmetries and the Masses of Gauge Bosons}",
    doi = "10.1103/PhysRevLett.13.508",
    journal = "Phys. Rev. Lett.",
    volume = "13",
    pages = "508--509",
    year = "1964"
}

@article{Englert:1964et,
    author = "Englert, F. and Brout, R.",
    editor = "Taylor, J. C.",
    title = "{Broken Symmetry and the Mass of Gauge Vector Mesons}",
    doi = "10.1103/PhysRevLett.13.321",
    journal = "Phys. Rev. Lett.",
    volume = "13",
    pages = "321--323",
    year = "1964"
}

@article{ATLAS:2012yve,
    author = "Aad, Georges and others",
    collaboration = "ATLAS",
    title = "{Observation of a new particle in the search for the Standard Model Higgs boson with the ATLAS detector at the LHC}",
    eprint = "1207.7214",
    archivePrefix = "arXiv",
    primaryClass = "hep-ex",
    reportNumber = "CERN-PH-EP-2012-218",
    doi = "10.1016/j.physletb.2012.08.020",
    journal = "Phys. Lett. B",
    volume = "716",
    pages = "1--29",
    year = "2012"
}

@article{CMS:2013btf,
    author = "Chatrchyan, Serguei and others",
    collaboration = "CMS",
    title = "{Observation of a New Boson with Mass Near 125 GeV in $pp$ Collisions at $\sqrt{s}$ = 7 and 8 TeV}",
    eprint = "1303.4571",
    archivePrefix = "arXiv",
    primaryClass = "hep-ex",
    reportNumber = "CMS-HIG-12-036, CERN-PH-EP-2013-035",
    doi = "10.1007/JHEP06(2013)081",
    journal = "JHEP",
    volume = "06",
    pages = "081",
    year = "2013"
}

@article{CEPCStudyGroup:2023quu,
    author = "Abdallah, Waleed and others",
    collaboration = "CEPC Study Group",
    title = "{CEPC Technical Design Report: Accelerator}",
    eprint = "2312.14363",
    archivePrefix = "arXiv",
    primaryClass = "physics.acc-ph",
    reportNumber = "IHEP-CEPC-DR-2023-01, IHEP-AC-2023-01",
    doi = "10.1007/s41605-024-00463-y",
    journal = "Radiat. Detect. Technol. Methods",
    volume = "8",
    number = "1",
    pages = "1--1105",
    year = "2024",
    note = "[Erratum: Radiat.Detect.Technol.Methods 9, 184--192 (2025)]"
}

@article{Adli:2025swq,
    author = "Adli, Erik and others",
    title = "{The Compact Linear e$^+$e$^-$ Collider (CLIC)}",
    eprint = "2503.24168",
    archivePrefix = "arXiv",
    primaryClass = "physics.acc-ph",
    month = "3",
    year = "2025"
}

@article{Arroyo-Urena:2024soo,
    author = "Arroyo-Ure{\~n}a, M. A. and Herrera-Chac{\'o}n, E. A. and Rosado-Navarro, S. and Salazar, Humberto",
    title = "{Hunting for a charged Higgs boson pair in proton-proton collisions}",
    eprint = "2405.06036",
    archivePrefix = "arXiv",
    primaryClass = "hep-ph",
    doi = "10.1103/PhysRevD.111.015023",
    journal = "Phys. Rev. D",
    volume = "111",
    number = "1",
    pages = "015023",
    year = "2025"
}

@article{HernandezSanchez:2012eg,
    author = "Hern\'andez-S\'anchez, J. and Moretti, S. and Noriega-Papaqui, R. and Rosado, A.",
    title = "Off-diagonal terms in Yukawa textures of the Type-III 2-Higgs doublet model and light charged Higgs boson phenomenology",
    journal = "JHEP",
    volume = "1307",
    pages = "044",
    year = "2013",
    doi = "10.1007/JHEP07(2013)044",
    archivePrefix = "arXiv",
    eprint = "1212.6818",
    primaryClass = "hep-ph"
}

@article{CMS:2017con,
    author = "Sirunyan, A. M. and others",
    collaboration = "CMS",
    title = "Search for lepton flavour violating decays of the Higgs boson to $\mu\tau$ and e$\tau$ in proton-proton collisions at $\sqrt{s}=13$ TeV",
    journal = "JHEP",
    volume = "06",
    pages = "001",
    year = "2018",
    doi = "10.1007/JHEP06(2018)001",
    archivePrefix = "arXiv",
    eprint = "1712.07173",
    primaryClass = "hep-ex"
}

@article{ATLAS:2019pmk,
    author = "Aad, G. and others",
    collaboration = "ATLAS",
    title = "Searches for lepton-flavour-violating decays of the Higgs boson in $\sqrt{s}=13$ TeV pp collisions with the ATLAS detector",
    journal = "Phys. Lett. B",
    volume = "800",
    pages = "135069",
    year = "2020",
    doi = "10.1016/j.physletb.2019.135069",
    archivePrefix = "arXiv",
    eprint = "1907.06131",
    primaryClass = "hep-ex"
}

@article{Workman:2022ynf,
    author = "Workman, R. L. and others",
    collaboration = "Particle Data Group",
    title = "Review of Particle Physics",
    journal = "PTEP",
    volume = "2022",
    pages = "083C01",
    year = "2022",
    doi = "10.1093/ptep/ptac097"
}

@article{Alloul:2013bka,
    author = "Alloul, A. and Christensen, N. D. and Degrande, C. and Duhr, C. and Fuks, B.",
    title = "FeynRules 2.0 - A complete toolbox for tree-level phenomenology",


    journal = "Comput. Phys. Commun.",
    volume = "185",
    pages = "2250-2300",
    year = "2014",
    doi = "10.1016/j.cpc.2014.04.012",
    archivePrefix = "arXiv",
    eprint = "1310.1921",
    primaryClass = "hep-ph"
}

@article{MadGraphNLO,
    author = "Alwall, J. and Herquet, M. and Maltoni, F. and Mattelaer, O. and Stelzer, T.",
    title = "MadGraph 5: going beyond",
    journal = "JHEP",
    volume = "06",
    pages = "128",
    year = "2011",
    doi = "10.1007/JHEP06(2011)128"
}

@techreport{Sjostrand:2008vc,
    author = "Sjostrand, T.",
    title = "PYTHIA 8 Status Report",
    institution = "DESY",
    year = "2009",
    doi = "10.3204/DESY-PROC-2009-02/41",
    archivePrefix = "arXiv",
    eprint = "0809.0303",
    primaryClass = "hep-ph"
}

@article{delphes,
    author = "de Favereau, J. and Delaere, C. and Demin, P. and Giammanco, A. and Lemaitre, V. and Mertens, A. and Selvaggi, M.",
    title = "DELPHES 3, A modular framework for fast simulation of a generic collider experiment",
    journal = "JHEP",
    volume = "02",
    pages = "057",
    year = "2014",
    doi = "10.1007/JHEP02(2014)057"
}

\end{document}